\documentclass[fleqn,10pt]{wlscirep}
\usepackage[utf8]{inputenc}
\usepackage[T1]{fontenc}
\usepackage{epsfig}
\usepackage{graphicx}
\usepackage{amsmath}
\usepackage{amssymb}
\usepackage{epstopdf}
\usepackage{xcolor}
\usepackage{subfigure}
\usepackage{bm}
\usepackage{dsfont}
\usepackage{mathrsfs}
\usepackage{hyperref}
\usepackage{todonotes}
\usepackage{comment}
\usepackage{physics}
\usepackage[normalem]{ulem}  % For \sout
\usepackage{soul}
\usepackage{verbatim}

\title{Persistence of Topological Phases in Non-Hermitian Quantum Walks}

\author[1]{Vikash Mittal}
\author[2]{Aswathy Raj}
\author[1]{Sanjib Dey}
\author[1,*]{Sandeep K.~Goyal}
\affil[1]{Department of Physical Sciences, Indian Institute of Science Education \& Research (IISER) Mohali, Sector 81 SAS Nagar, Manauli PO 140306 Punjab, India}
\affil[2]{Department of Physics, Indian Institute of Science Education \&  Research (IISER) Bhopal, Bhopal Bypass Road, Bhauri, Bhopal 462066, India}

\affil[*]{skgoyal@iisermohali.ac.in}

\begin{abstract}
	Discrete-time quantum walks are known to exhibit exotic topological states and phases. Physical realization of quantum walks in a lossy environment may destroy these phases. We investigate the behaviour of topological states in quantum walks in the presence of a lossy environment. The environmental effects in the quantum walk dynamics are addressed using the non-Hermitian Hamiltonian approach. We show that the topological phases of the quantum walks are robust against moderate losses. The topological order in one-dimensional split-step quantum walk persists as long as the Hamiltonian respects exact $\mathcal{PT}$-symmetry. Although the topological nature persists in two-dimensional quantum walks as well, the $\mathcal{PT}$-symmetry has no role to play there. Furthermore, we observe  topological phase transition in two-dimensional quantum walks that is induced by losses in the system.
\end{abstract}
\begin{document}

\flushbottom
\maketitle

\thispagestyle{empty}

\section{Introduction}
Quantum walks are the quantum analogue of classical random walks\cite{Aharonov1993, Ambainis2001, Kempe2003, VenegasAndraca2012, Nayak2000} where a quantum walker propagates on a lattice and the direction of propagation is conditioned over the state of its coin. Due to the quantum nature of the walker and the coin, the position state of the walker is a superposition of multiple lattice sites. This provides a quadratically fast spread of the walker over the lattice as compared to its classical counterpart \cite{Ambainis2001}. Quantum walks, continuous-time as well as discrete-time, are important in various fields including universal quantum computation \cite{Childs2009, Childs2013, Lovett2010}, quantum search algorithms \cite{Ambainis2003, Childs2004, Shenvi2003, Agliari2010}, quantum simulations \cite{Nicola2014}, quantum state transfer \cite{Wojcik2011} and simulation of physical systems \cite{Schreiber2012, Sansoni2012, Peruzzo2010}. Quantum walks have been used in other branches of science as well, such as in biology to study the energy transfer in photosynthesis \cite{Mohseni2008}. They have also been proved as a promising candidate to simulate the decoherence \cite{Romanelli2005, Kendon2007} and to implement generalized measurements (POVM) \cite{Wojcik2013}.

Quantum walks have started gaining popularity among condensed matter physicists since the last decade because one can simulate exotic topological phases using one (1D) and two-dimensional (2D) discrete-time quantum walk (DTQW)~\cite{Kitagawa2010, Kitagawa2012, Asboth2012, Asboth2015}. As a consequence, people have been able to establish the bulk-boundary correspondence for 1D periodic systems \cite{Asboth2013, Asboth2014}. The versatile nature of quantum walks make them a prime candidate for fault-tolerant topological quantum computation and quantum simulations. 

Quantum walks have been implemented on a variety of systems, such as; trapped ions/atoms \cite{Milburn2002, Schmitz2009, Roos2010, Karski2009}, optical systems \cite{Schreiber2010, Schreiber2011, Regensburger2012, Broome2010, Zhang2007, Sephton2019}, NMR \cite{Du2003, Laflamme2005}, Bose-Einstein condensate \cite{Sandro2017}, etc.
However, no quantum system is without losses, due to which implementation of quantum algorithms as well as the observation of exotic topological phases have always been difficult. In this article, we study the effect of losses on the topological phases arising in quantum walk systems. A system along with losses effectively renders the quantum walk dynamics non-unitary. We treat this non-unitary evolution using the non-Hermitian Hamiltonian approach \cite{Bender1998,Mostafazadeh2002}. We establish that for a 1D split-step quantum walk (SSQW), the topological phases persist as long as the spectrum of the non-Hermitian Hamiltonian is real. In other words, as long as the underlying non-Hermitian Hamiltonian is exact $\mathcal{PT}$-symmetric \cite{Bender1998}, the topological phase is preserved. 2D quantum walks have a more complex structure. In this case, too, we observe the persistence of the topological phases. However,  $\mathcal{PT}$ symmetry is absent in 2D DTQW and the quasi-energies are complex even for infinitesimal losses. Furthermore, loss-induced topological phase transition can be observed in these 2D DTQWs. We study the bulk-boundary correspondence to reconfirm our results and numerically show the robustness of the edge states with the introduction of losses.

The non-Hermitian quantum walk has been studied theoretically \cite{Mochizuki2016} as well as experimentally~\cite{Regensburger2012}. The existence of topological edge states \cite{PengXue2017}, topological phase transition \cite{Rudner2009,Zhang2017,Wang2019} as well as the correspondence between bulk and boundary in non-Hermitian quantum walks have also been established \cite{PengXue2020}. In \cite{Sanders2017}, the authors have introduced the non-Hermiticity by making partial measurements on the internal states of the walker and showed the robustness of the topological phases against the disorder. The same model was extended to study higher winding numbers \cite{PengXue2018}. The role of time-reversal symmetry in topologically protected states was studied in \cite{McGinley2020}. We use a different model to introduce non-Hermiticity and establish the \textit{persistent} nature of topological phases in these systems. We show that the topological nature of the underlying Hamiltonian does not change in the lossy environment within certain limits.  We show that in 1D SSQW the topological phase persists as long the system possess exact $\mathcal{PT}$-symmetry. In the case of a 2D quantum walk, such correspondence between the topological order and the $\mathcal{PT}$-symmetry is missing. In this systems, an  interesting observation is the loss-induced topological phase transition, which is absent in the 1D case.

The article is organized as follows: In Sec.~\ref{Sec:Background}, we discuss the topics which are relevant for the understanding of our results. Sec.~\ref{Sec:Results} contains our results on the effect of the losses on the topological nature of quantum walks. Here, we discuss the 1D SSQW and 2D quantum walks and show the persistence of topological phases in a noisy environment. We conclude in Sec.~\ref{Sec:Conclusion}.

\section{Background} \label{Sec:Background}
In this section, we introduce the topics which are relevant to understand our results. We start with 1D and 2D unitary as well as non-unitary DTQWs. Specifically, we discuss the 1D and 2D DTQW, and 1D SSQW and the topological classes arising in these systems. Methods to characterize the topological phases are also discussed in this section.

\subsection{1D DTQW} \label{Sec:1D DTQW}
DTQW of a quantum walker over a one-dimensional lattice consists of a conditional shift operator $T$ and a  coin flip operator $R(\theta)$ for a real parameter $\theta$. In position basis $\{\ket{n}\} \in \mathcal{H}_{\text{pos}}$ and spin basis $\{\ket{\uparrow}, \ket{\downarrow} \}$, the operator $U(\theta) = T R(\theta)$ governs the time evolution of the walker for a unit time on the lattice. Here 
\begin{align}
	&T = \sum_n \ket{\uparrow} \bra{\uparrow} \otimes \ket{n+1} \bra{n}  + \ket{\downarrow} \bra{\downarrow} \otimes \ket{n-1} \bra{n} ,\label{eq:1D-Translation} \\
	&R(\theta) = e^{-i \theta \sigma_y/2} \otimes \mathds{1},
\end{align}
and $-2\pi\le\theta<2\pi$ is a real parameter and $\sigma_y$ is the Pauli matrix along the $y$-axis. Here, $\mathds{1}$ represents the identity operation on the lattice. The operator $U(\theta)$ can be expressed in terms of the underlying Hamiltonian $H(\theta)$ as $U(\theta) = e^{-i H(\theta)}$~\cite{Kitagawa2010}. For simplicity, we have assumed $\hbar = 1$ and the periodic boundary condition with $N$ number of lattice sites. Since the unitary operator $U(\theta)$ and the Hamiltonian is translation invariant, the (quasi) momentum eigenbasis 
$\{\ket{k}\}$ are also the energy eigenstates. These states are defined as
\begin{equation*}
	\ket{n} = \dfrac{1}{\sqrt{N}} \sum_{k}  \omega^{k n} \ket{k},\quad \omega = e^{i 2 \pi/N},
\end{equation*}
with $-\pi \le k \le \pi$ being the quasi-momentum. The Hamiltonian $H(\theta)$ in the quasi-momentum space reads~\cite{Kitagawa2010}
\begin{equation} \label{eq:1D-Hamil}
	H(\theta) = \sum_{k} [E_{\theta}(k)\, \vb{n}_{\theta}(k)\vdot \vb{\sigma}] \otimes \dyad{k}, 
\end{equation}
where the energy $E_{\theta}(k)$ and  the unit Bloch vector $\vb{n}_{\theta}(k)$  read $\cos E_{\theta}(k) = \cos (\theta/2) \cos k $, and 
\begin{align}
	\vb{n}_{\theta}(k) &= \dfrac{(\sin(\theta/2) \sin k, \sin(\theta/2) \cos k, -\cos(\theta/2) \sin k)}{\sin E_{\theta}(k)}.
\end{align}
\subsection{1D SSQW} \label{Sec:1D SSQW}
A more enriched class of 1D DTQW is SSQW, which involves splitting the conditional shift operator $T$ into left-shift ($ T_{\downarrow} $) and right-shift ($ T_{\uparrow} $) operators, separated by an additional coin toss $R(\theta_2)$~\cite{Kitagawa2010}. The resultant time evolution operator for split-step quantum walks (in one-dimension) reads
\begin{equation} \label{eq:Unitary-SSQW}
	U_{_{\text{SS}}}(\theta_1, \theta_2) = T_{\downarrow} R(\theta_2) T_{\uparrow} R(\theta_1),
\end{equation}
where
\begin{align*}
	T_{\downarrow} &= \sum \ket{\uparrow} \bra{\uparrow} \otimes \mathds{1}  + \ket{\downarrow} \bra{\downarrow} \otimes \ket{n-1} \bra{n}, \\
	T_{\uparrow} &= \sum \ket{\uparrow} \bra{\uparrow} \otimes \ket{n+1} \bra{n}  + \ket{\downarrow} \bra{\downarrow} \otimes \mathds{1}.
\end{align*}
In this case, the effective Hamiltonian $H_{\text{SS}}(\theta_1, \theta_2)$ can be written down in quasi-momentum space as
\begin{equation} \label{eq:SSQW-Hamil}
	H_{_{\text{SS}}}(\theta_1, \theta_2) = \sum_{k} [E_{\theta_1, \theta_2}(k) \vb{n}_{\theta_1, \theta_2}(k)\vdot \vb{\sigma}] \otimes \dyad{k}.
\end{equation}
The energy and the components of the Bloch vector are given by 
\begin{align}
	\cos E_{\theta_1, \theta_2}(k) &= \cos(\theta_1/2) \cos(\theta_2/2) \cos k  -\sin(\theta_1/2) \sin(\theta_2/2),
\end{align}
and $\vb{n}_{\theta_1, \theta_2}(k) = n_x(k) \vb{\hat{i}} + n_y(k) \vb{\hat{j}} + n_z(k) \vb{\hat{k}}$ with
\begin{align}
	n_x(k) &= \dfrac{\sin(\theta_1/2)\cos(\theta_2/2)\sin k}{\sin E_{\theta_1, \theta_2}(k)}, \nonumber \\
	n_y(k) &= \dfrac{\cos(\theta_1/2) \sin(\theta_2/2) +   \sin(\theta_1/2) \cos(\theta_2/2) \cos k }{\sin E_{\theta_1, \theta_2}(k)},\nonumber \\
	n_z(k) &= \dfrac{-\cos(\theta_1/2)\cos(\theta_2/2) \sin k}{\sin E_{\theta_1, \theta_2}(k)} \nonumber.
\end{align}
Even though 1D SSQW seems complicated when it comes to implementation, it is not much different from ordinary 1D DTQW; Mathematically, the 1D SSQW  can be decomposed in two steps of ordinary 1D DTQW~\cite{Zhang2017} and the time evolution operator of 1D SSQW can be written as
\begin{equation} \label{eq: Decom-SSQW}
	U_{_{\text{SS}}}(\theta_1, \theta_2) = U(\theta_2)U(\theta_{1}),
\end{equation}
where $U(\theta_i)$ is the time evolution for 1D DTQW.
\subsection{2D DTQW}  \label{Sec:2D DTQW}
\begin{figure}
	\centering
	\includegraphics[width=6cm]{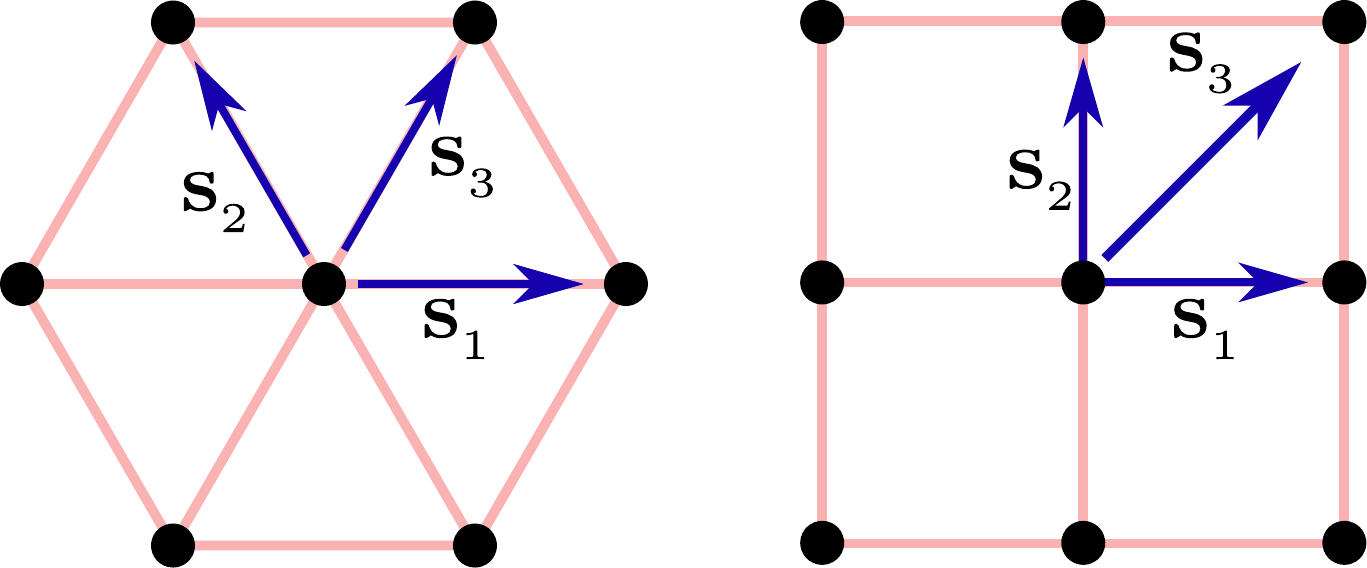}
	\caption{(Color online) 2D DTQW with nontrivial topology on a triangular lattice and its equivalent square lattice.}
	\label{fig:tri}
\end{figure}

There are several ways of defining a 2D DTQW in a lattice. For our purpose, we introduce the one in which we have a square lattice and a two-dimensional coin~\cite{Kitagawa2010}. This DTQW consists of two conditional translations in two directions accompanied by rotation of the coin. The time evolution operator of 2D DTQW can be written as
\begin{equation} \label{eq:2D-Unitary}
	U^{'}_{_{2D}}(\theta_1, \theta_2) = T_y R(\theta_2) T_x R(\theta_1), 
\end{equation}
where $ T_x $ and $ T_y $ are the translation operators,  which translate the particle in  $x$ and  $y$ directions, respectively. We can also define, 2D DTQW on a triangular lattice which consists of three spin-dependent translations separated by coin-flip operations. In that case, the unitary operator which governs the time evolution is written as
\begin{equation} \label{eq:U2D}
	\tilde U_{_{2D}}(\theta_1, \theta_2) = T_{xy} R(\theta_1) T_y R(\theta_2) T_x R(\theta_1),
\end{equation}
where $ T_i (i=x,y,xy) $ are the translations along $ \vb{s}_i  $ directions with $ T_{xy} = T_xT_y $, as shown in Fig.\,\ref{fig:tri}. We can derive another two-dimensional quantum walk which is unitarily equivalent to $\tilde U_{_{2D}}(\theta_1, \theta_2)$ as $\tilde U_{_{2D}} \to U_{_{2D}} = T_x^\dagger \tilde U_{_{2D}} T_x$. The resulting time evolution unitary operator can be written as
\begin{equation}
	U_{_{2D}}(\theta_1, \theta_2) = T_y R(\theta_1) T_y R(\theta_2) T_x R(\theta_1) T_x.\label{Eq:Qwalk2D}
\end{equation}
The underlying  Hamiltonian for this quantum walk (in quasi-momentum space) reads
\begin{equation} \label{eq:Hamil2D}
	H_{_{2D}}(\theta_1, \theta_2) = \sum_{k_x, k_y} E(k_x,k_y) \hat{\vb{n}}(k_x,k_y) \vdot \vb{\sigma} \otimes \dyad{k_x, k_y},
\end{equation}
\noindent where the expression of quasi-energy reads
\begin{align}
	\cos E(k_x, k_y) &=  \cos \theta_1 \cos(\theta_2/2) \cos^2(k_x + k_y)  - \sin \theta_1\sin(\theta_2/2)\cos(k_x + k_y)\cos(k_x - k_y) \nonumber\\
	&\;  -\cos(\theta_2/2) \sin^2(k_x + k_y) ,
\end{align}
and the Bloch vector reads \cite{Kitagawa2010}
\begin{equation*}
	\hat{\vb{n}}(k_x,k_y) = \dfrac{n_x(k_x,k_y) \hat{\vb{i}} + n_y(k_x,k_y) \hat{\vb{j}} + n_z(k_x,k_y) \hat{\vb{k}}}{\sin E(k_x, k_y)},
\end{equation*}
with
\begin{align}
	n_x(k_x,k_y) = &- \sin \theta_1\cos(\theta_2/2) \cos(k_x + k_y) \sin (k_x - k_y) - \cos^2 \theta_1 \sin(\theta_2/2) \sin 2(k_x - k_y), \nonumber \\
	n_y(k_x,k_y) = &\sin \theta_1 \cos(\theta_2/2) \cos(k_x + k_y) \cos(k_x - k_y) + \cos \theta_1  \cos^2(k_x - k_y) \sin(\theta_2/2) - \sin^2(k_x - k_y) \sin(\theta_2/2),\nonumber \\
	n_z(k_x,k_y) = &-\cos^2(\theta_1/2) \cos(\theta_2/2) \sin 2(k_x + k_y) + \sin \theta_1\sin(\theta_2/2)\sin(k_x + k_y)\cos(k_x - k_y) \nonumber.
\end{align}
The purpose of writing the evolution for 2D DTQW as \eqref{Eq:Qwalk2D} is that now it can be decomposed as  two 1D SSQW in two different directions, i.e.,~\cite{Zhang2017} 
\begin{equation}
	U_{_{2D}}(\theta_1, \theta_2) = U^y_{_{\text{SS}}}(\theta_1, 0) U^x_{_{\text{SS}}}(\theta_1, \theta_2),
\end{equation}
where $ U^i_{_{\text{SS}}} $ is the time-evolution operator of 1D SSQW \eqref{eq:Unitary-SSQW}.

\subsection{Characterizing Topological Phases}\label{Sec2d}
% Topological phase or topological order is defined as the ground state degeneracy in a quantum system due to the topological properties of the parameter space.
The class of topological phases which can be realized in a system is characterized by the underlying symmetries of the Hamiltonian and the dimensionality of the system. They are further quantified by nonlocal topological invariants and possess non-Abelian geometric phases which are quantized \cite{Bernevig2013, Chiu2016, Shankar2018}. The topological and nonlocal nature of these phases make them robust against local perturbations.  Tuning the parameters of the Hamiltonian may result in the system going from one topological phase to another as a result of topological phase transition, without breaking the underlying symmetry of the Hamiltonian. We will discuss the symmetries of the Hamiltonian later in detail.

Topological phases can be characterized and classified into various classes using different parameters. In 1D systems, the winding number is the topological invariant that characterizes the topological phase. For a given  Hamiltonian $H = \bigoplus_k H(k)$, the winding number $W_m$ for the $ m $th band is defined as
\begin{equation} \label{eq: Winding-Number}
	W_m = \dfrac{1}{\pi} \int_{\Gamma} \mathcal{A}_m(k) dk,
\end{equation}
where $\mathcal{A}_m$ is the Berry connection given as~\cite{Berry1984}
\begin{equation*}
	\mathcal{A}_m(k) = -i  \expval{\dfrac{\partial}{\partial k}}{\psi_m(k)}.
\end{equation*}
Here, $\ket{\psi_m(k)} $ is the $ m $th eigenstates of $H(k)$ for the parameter value $ k $. By definition, the Winding number $W$ is the Berry phase divided by $\pi$. In the case of discrete system, it would be convenient to calculate the Berry phase using Pancharatnam's connection \cite{Mukunda1993}. For a given set of pure states 
$\{\ket{\psi_m(k_n)}\}$ for $m$th band, where $k_n$ is momentum (quasi) which takes discrete values, it reads
\begin{equation}
	\gamma_m = -\arg{\ip*{\psi_m(k_1)}{\psi_m(k_2)}\ip*{\psi_m(k_2)}{\psi_m(k_3)}\ip*{\psi_m(k_3)}{\psi_m(k_4)} \dots \ip*{\psi_m(k_N)}{\psi_m(k_1)} }.
\end{equation}
Geometrically, the winding number $W_m$ represents the number of times the Bloch vector $\vb{n}$ corresponding to the state $\ket{\psi_m(k)}$ goes around the origin in the counter-clockwise direction as $k$ runs over the first Brillouin Zone. Note that winding numbers are sufficient to characterize the topological order in translation-invariant systems~\cite{Cedzich2018}. However, in more complex systems such as, systems with disorder in the coin angles or systems in which two bulks are connected through some crossover region, additional invariants may be required for such characterization~\cite{Cedzich2016, Cedzich2018b}. They can be calculated easily using Schur approach~\cite{Cedzich2019}.

In two or higher dimensional systems, Chern number \cite{Asboth2016} is one of the topological invariants which is used and defined as
\begin{equation} \label{eq:Chern-Number}
	C_m = \dfrac{1}{2 \pi} \oint_S \mathcal{F}_m d^2 \vb{k},
\end{equation}
for  $ \mathcal{F}_m = \nabla \times \mathcal{A}_m $ and the integration is over the closed surface in two-dimensions (two-dimensional Brillouin zone). Here, $\mathcal{A}$ is the Berry curvature. 

\begin{figure}
	\centering
	\subfigure[]{
		\includegraphics[width=4cm]{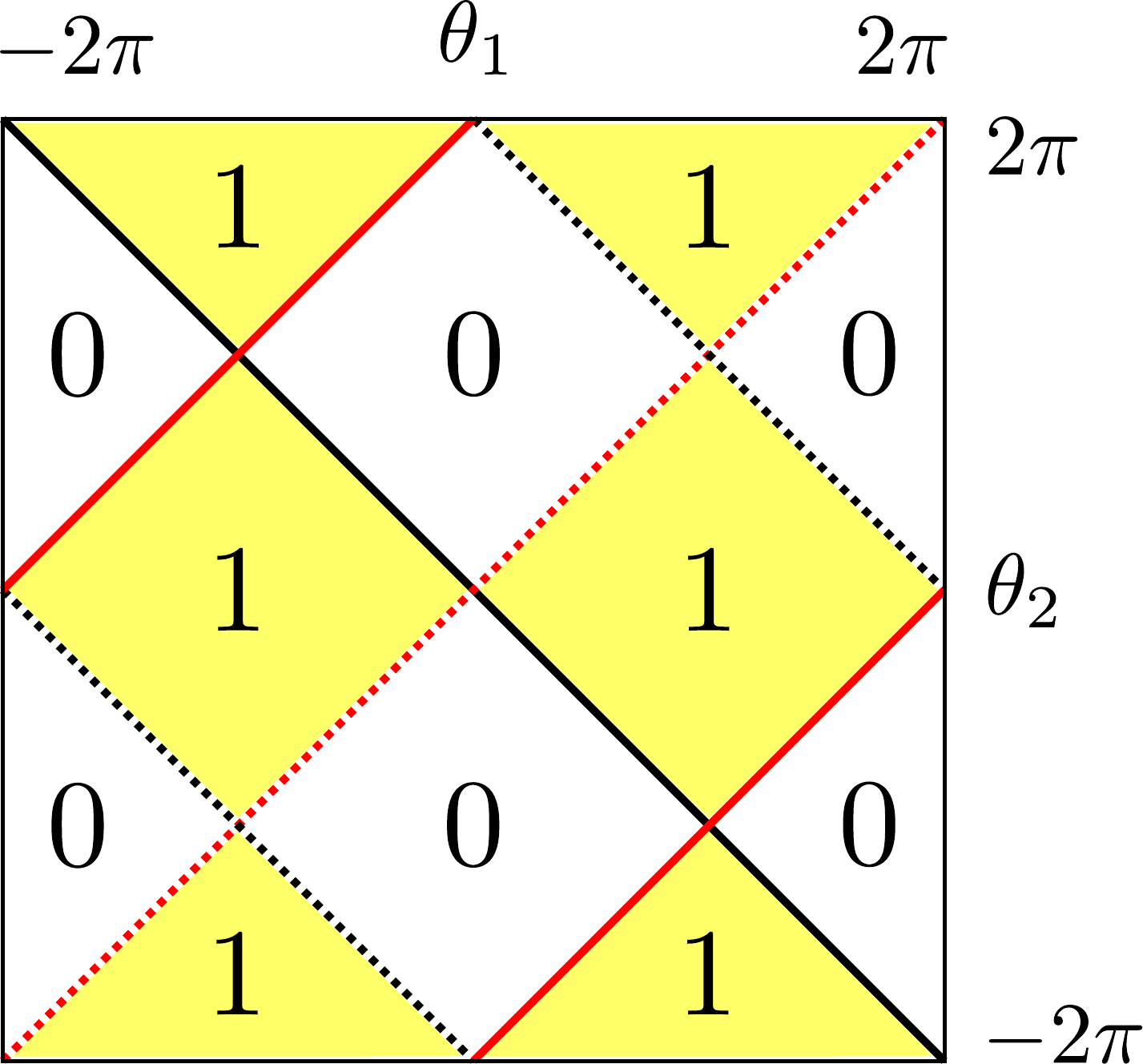}
		\label{fig:PSQW}}
	\subfigure[]{
		\includegraphics[width=4cm]{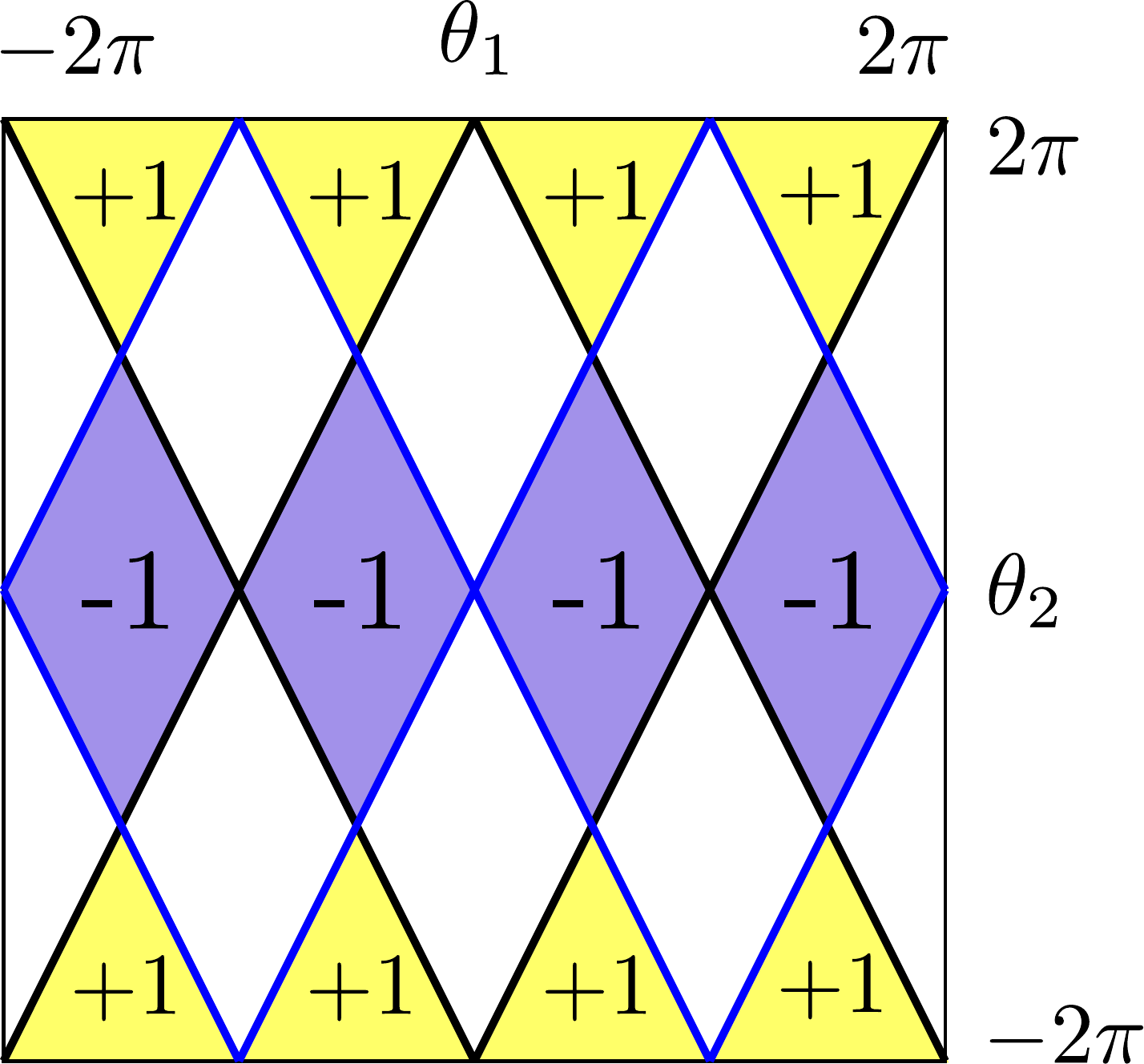}
		\label{fig:PS2DQW}}
	\caption{(Color online) \subref{fig:PSQW} Different topological phases realized in 1D SSQW as a function of $\theta_1$ and $\theta_2$. We observe two topological phases here corresponding to $W = 0$ and $1$. Here, black and red lines represent closing of energy band at $k = 0$ and $k = \pi$, respectively, and solid and dotted lines demonstrate the closing at $E = 0$ and $E = \pi$, respectively. \subref{fig:PS2DQW} Topological phases which exist in 2D DTQW for different values of $\theta_1$ and $\theta_2$. Here, blue and black lines show the closing of energy gap at $ E = 0 $ and $E = \pi$, respectively. The yellow, violet and white regions correspond to $ C = +1, -1 $ and $ 0 $, respectively.}
	\label{fig:PS}
\end{figure}

Quantum walk Hamiltonian possesses a rich topological structure. For example, the Hamiltonian $H_{_{\text{SS}}}(\theta_1, \theta_2)$ \eqref{eq:SSQW-Hamil} corresponding to 1D SSQW  with parameters $\theta_1$ and $\theta_2$ exhibits two different topological phases characterized by the winding number $W = 0$ and $W = 1$, as shown in Fig.~\ref{fig:PSQW}~\cite{Kitagawa2010}. In Fig.~\ref{fig:PS2DQW}, we plot the  topological phases with Chern number $ C = 0, \pm 1 $ exhibited by the Hamiltonian  $H_{_{2D}}$ \eqref{eq:Hamil2D} for 2D DTQW~\cite{Kitagawa2010}.

\subsection{Non-unitary Quantum Walk}
Generally, quantum walk dynamics is given by a unitary time evolution operator. However, limitations in physical implementation and the environmental effects can cause losses which can cause the dynamics to deviate from unitary nature. 
In general, one can extend 1D SSQW to a non-unitary quantum walk by introducing a scaling operator $G$~\cite{Mochizuki2016}, with tunable parameters in the dynamics. The resulting time evolution operator for a non-unitary quantum walk can be written as 
\begin{equation} \label{eq: Unitary NonUSSQW}
	U^{^{\text{NU}}}_{_{\text{SS}}} = T_{\downarrow} G_2 R(\theta_2) T_{\uparrow} G_1 R(\theta_1),
\end{equation}    
with
\begin{align}
	G_i &= \sum_{n} \begin{pmatrix}
		g_{i, \uparrow} (n) & 0 \\
		0  & g_{i, \downarrow} (n)
	\end{pmatrix} \otimes \dyad{n}.
\end{align}
If $g_{i, \uparrow},~g_{i, \downarrow} \ne$ 1 then $G_i$, as well as, $U$ become nonunitary. For simplicity, we consider the case of the homogeneous quantum walk, where all the $g_i(n)$s are independent of $n$ and the scaling operator is written as
\begin{align}
	G_2 &= G_1^{-1} = G_{\delta} = \begin{pmatrix}
		e^{\delta} & 0 \\
		0 & e^{-\delta}
	\end{pmatrix} \otimes \mathds{1}. 
\end{align} 
The above choice of operators is motivated by the experimental setup used in~\cite{Regensburger2012}. The factor $\delta$ is known as the loss and gain factor as the operator $G$ results in increasing (decreasing) the amplitude of spin-up (down). The time evolution operator for the non-unitary quantum walk becomes 
\begin{equation} \label{eq:SSQW-TimeEvolution}
	U^{^{\text{NU}}}_{_{\text{SS}}} = T_{\downarrow} G_{\delta} R(\theta_2) T_{\uparrow} G_{\delta}^{-1} R(\theta_1).
\end{equation}
This particular choice of the scaling operator leaves the translational symmetry of the quantum walk intact. Hence, the dynamical operator can be block-diagonalized in the momentum basis as
\begin{equation} \label{Uk}
	U^{^{\text{NU}}}_{_{\text{SS}}} = \sum_{k} \tilde{U}^{^{\text{NU}}}_{_{\text{SS}}}(k) \otimes \dyad{k}, 
\end{equation}
where
\begin{equation} \label{eq:NonUnitary-SSQW}
	\tilde{U}^{^{\text{NU}}}_{_{\text{SS}}}(k) = T_{\downarrow}(k) G_{\delta} R(\theta_2) T_{\uparrow}(k) G_{\delta}^{-1} R(\theta_1),
\end{equation}      
with $ T_{\downarrow}(k) = e^{i k (\sigma_z - \mathds{1})/2} $, $ T_{\uparrow}(k) = e^{i k (\sigma_z + \mathds{1})/2} $ and it acts only on the coin part. The corresponding generator or an effective Hamiltonian $H_{_{\text{NU}}}(\theta_1, \theta_2,  \delta)$ reads
\begin{equation} \label{eq:Hamil-SSQW}
	H_{_{\text{NU}}}(\theta_1, \theta_2, \delta) = \bigoplus_k  E(k)\, \hat{\vb{n}}(k) \vdot \vb{\sigma},
\end{equation}
with quasi-energy
\begin{equation} \label{eq: Energy-SSQWL}
	\cos E(k) = \cos (\theta_1/2) \cos (\theta_2/2) \cos k - \sin (\theta_1/2) \sin (\theta_2/2) \cosh 2 \delta,
\end{equation}
and $\vb{\hat{n}} = n_x(k) \vb{\hat{i}} + n_y(k) \vb{\hat{j}} + n_z(k) \vb{\hat{k}}$ with
\begin{align}
	n_x(k) &= \dfrac{\sin(\theta_1/2) \cos(\theta_2/2) \sin k  - i \cos(\theta_1/2) \sin(\theta_2/2) \sinh 2\delta}{\sin E(k)}, \nonumber \\
	n_y(k) &= \dfrac{\sin(\theta_1/2) \cos(\theta_2/2) \cos k + \cos(\theta_1/2) \sin(\theta_2/2) \cosh 2\delta}{\sin E(k)}, \nonumber \\
	n_z(k) &= \dfrac{- \cos(\theta_1/2) \cos(\theta_2/2) \sin k - i \sin(\theta_1/2) \sin(\theta_2/2) \sinh 2\delta}{\sin E(k)}.
\end{align}
Note that, for $\delta \ne 0$, $ G $ and $ U^{^{\text{NU}}}_{_{\text{SS}}} $ are no longer unitary operators and the norm of the state in the evolution may  not be preserved. Consequently, $H_{_{\text{NU}}}(\theta_1, \theta_2, \delta)$ is not Hermitian but still we have a real spectrum up to a certain critical value of $\delta = \delta_c$. Given the fact that the energy band closes at $k = 0, E = 0$ and from  \eqref{eq: Energy-SSQWL} we have an expression for $\delta_c$ which reads
\begin{equation} \label{Eq:Delta-c}
	\delta_c = \dfrac{1}{2}\cosh^{-1}\left[\dfrac{\cos (\theta_1/2) \cos (\theta_2/2)  - 1}{\sin (\theta_1/2) \sin (\theta_2/2)}\right].
\end{equation}
The argument of $\cosh^{-1}$ in the above equation is positive (or negative) when $\theta_1$ and $\theta_2$ have the opposite (or same) sign. The negative argument results in complex value of $\delta_c$. So we consider a complex form of  $ \delta $ given by $\delta = \gamma + i \phi$.
We observe that the negative argument of $\cosh^{-1}$ results in  $\phi_c = \pi/2$. In this article, we restrict ourselves to the case when $\delta_c$ is real, i.e., opposite signs for $\theta_1$ and $\theta_2$ and refer $\gamma$ as the scaling factor. The calculations for the case when $\phi_c = \pi/2$ are exactly the same as for $\phi_c = 0$ case. The imaginary value $\phi_c = \pi/2$ amounts to shifting $k \to k+\pi/2$. The $\gamma_c (= \delta_c)$ is the point where the exact $\mathcal{PT}$-symmetry (will be discussed in the next section) of the system breaks spontaneously (also known as the exceptional point \cite{Ozdemir2019}), and we will have complex energies for $\gamma > \gamma_c$. 

\begin{figure*}
	\centering
	\subfigure[]{
		\includegraphics[width=5cm]{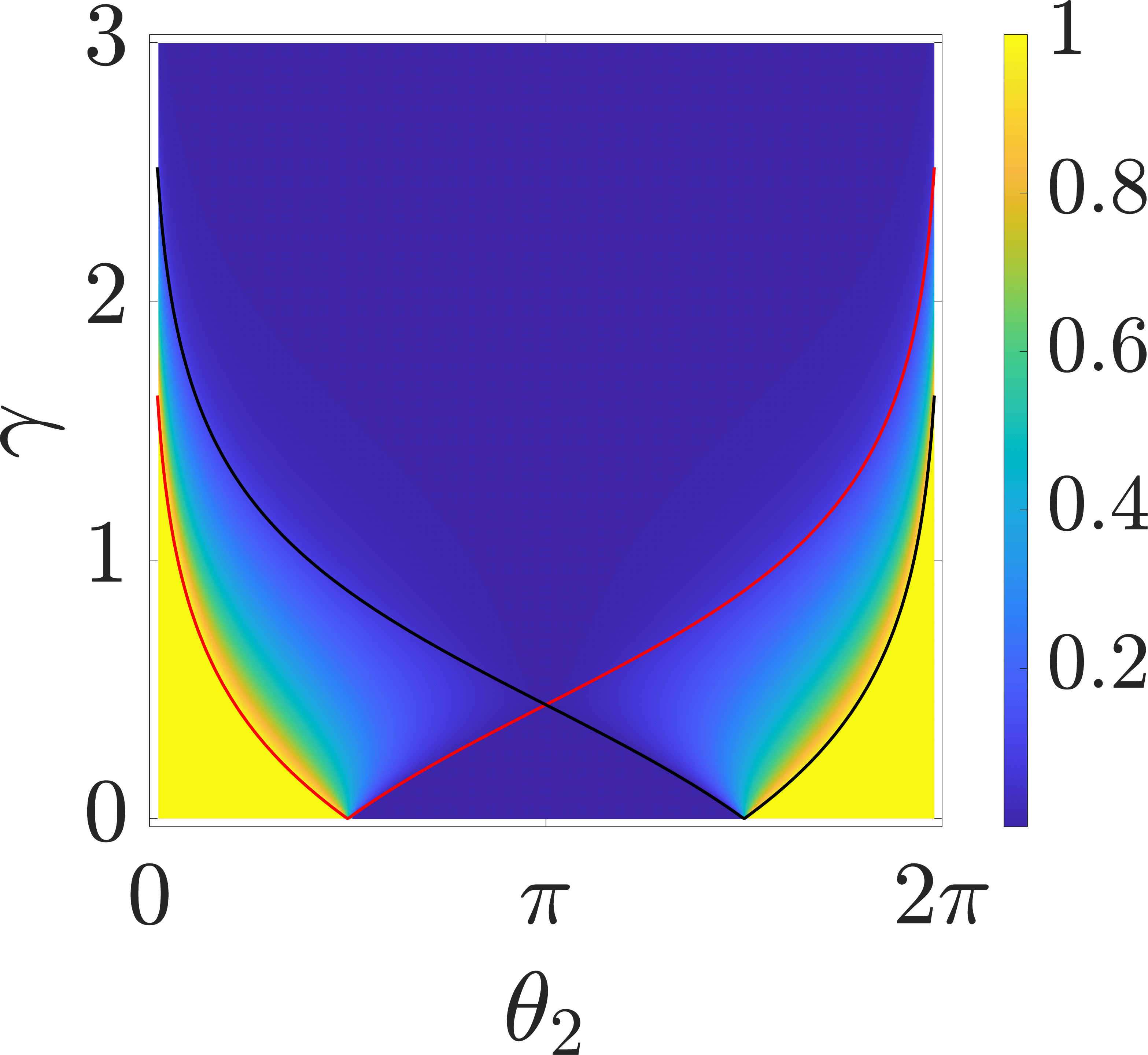}
		\label{fig:SSQWL1}}
	\subfigure[]{
		\includegraphics[width=5cm]{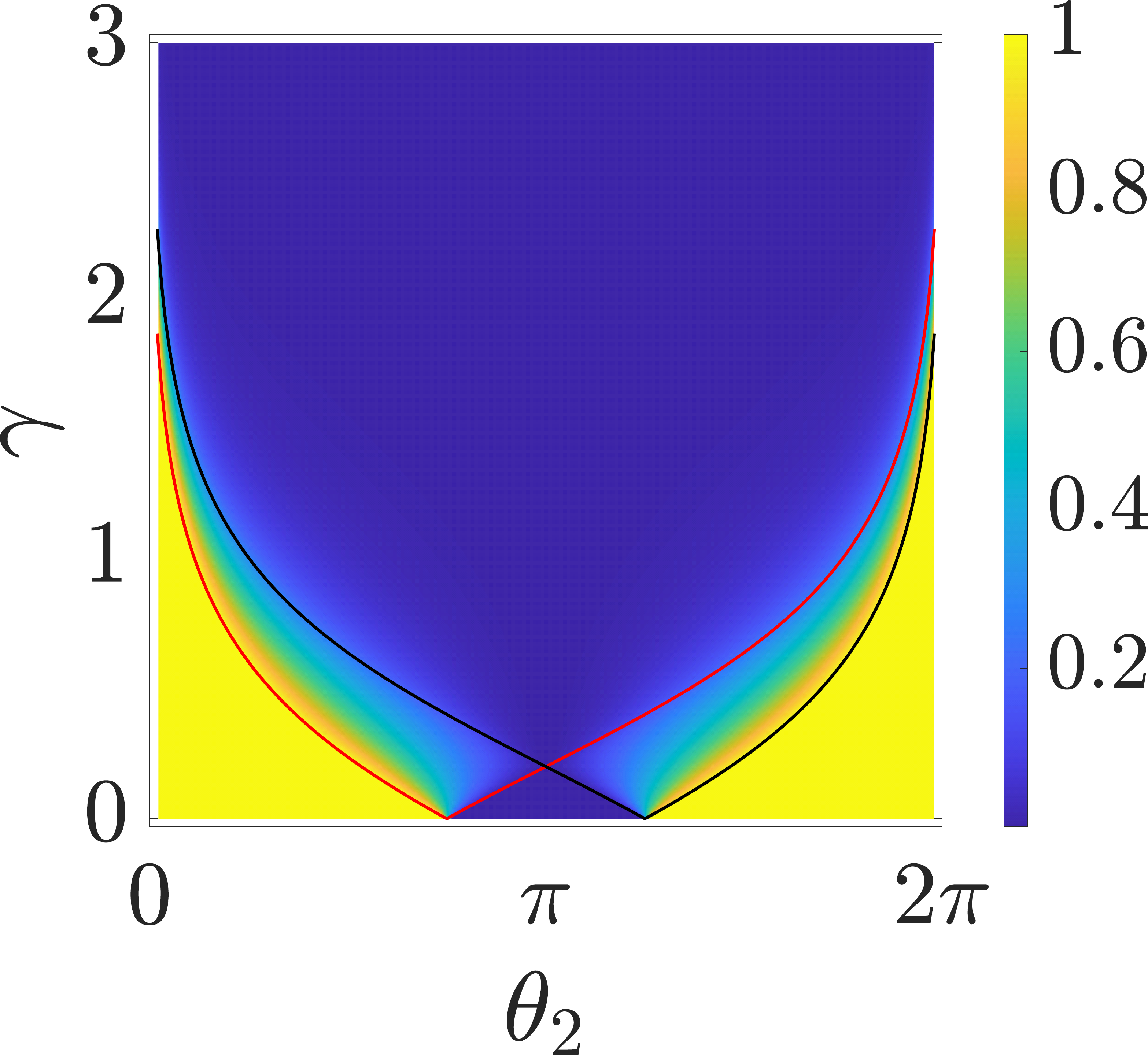}
		\label{fig:SSQWL2}}
	\subfigure[]{
		\includegraphics[width=5cm]{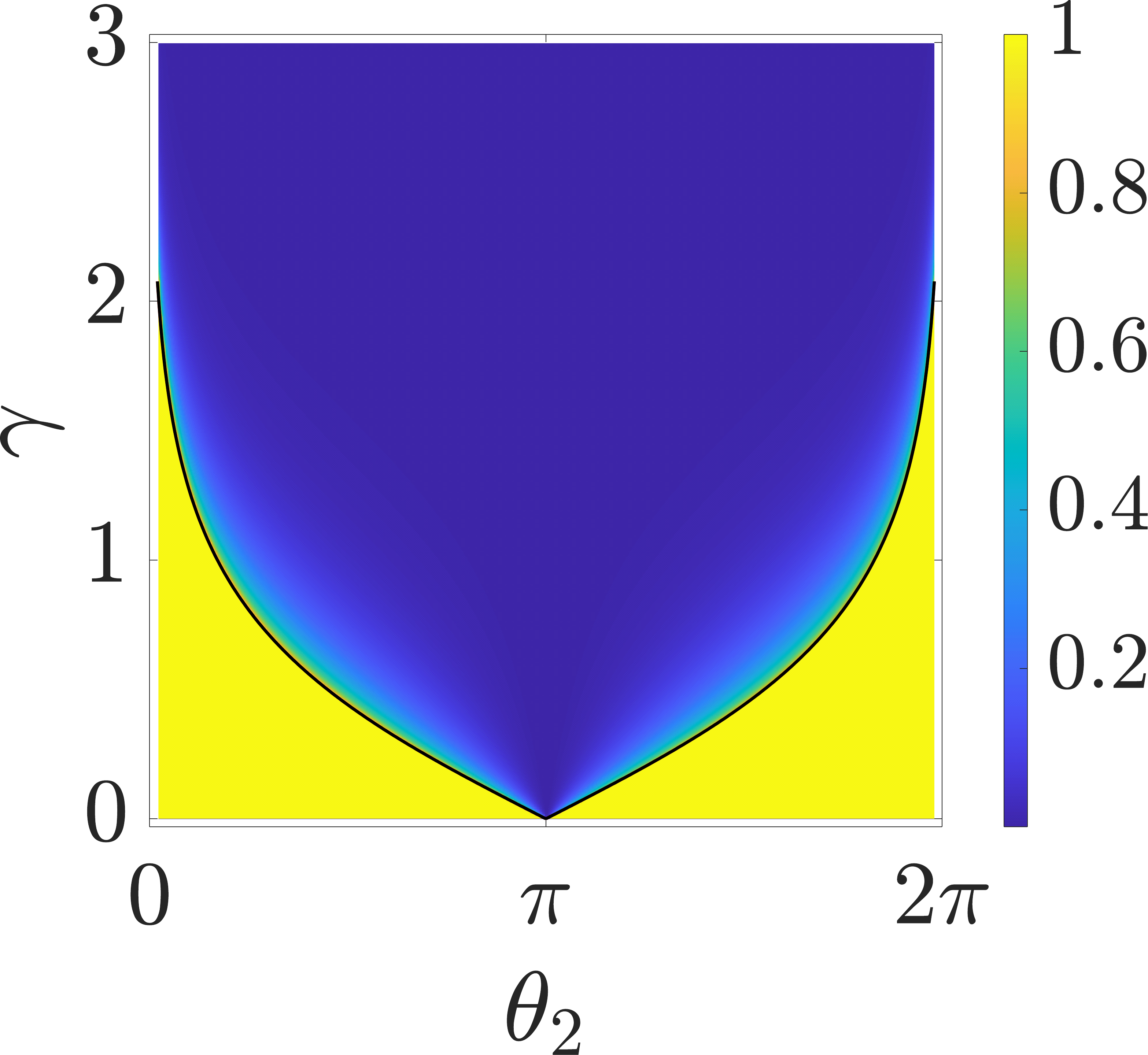}
		\label{fig:SSQWL3}}
	\caption{(Color online) Plot for $W_-$  for lower energy band  as a function of $\gamma$ and $\theta_2$, and \subref{fig:SSQWL1} $\theta_1 = -\pi/2$ \subref{fig:SSQWL2} $\theta_1 = -3 \pi/4$ \subref{fig:SSQWL3} $\theta_1 = -\pi$. The system size is taken to be $N = 201$. The red and black lines in all of the panels represent $\gamma_c$ for $(k,E)=(0,0)$ and $(k,E)=(\pi,0)$, respectively.}
	\label{fig:SSQWL}
\end{figure*}

\subsection{Symmetries of the Hamiltonian}
In this section, we discuss various  symmetries possessed by non-unitary quantum walks under consideration. We focus mainly on the parity and time reversal ($\mathcal{PT}$) symmetry, particle-hole ($\Xi$) symmetry  and the Chiral ($\Gamma$) symmetry. The $\mathcal{PT}$-symmetry characterizes the spectrum of the non-Hermitian Hamiltonian whereas the particle-hole and the Chiral symmetry classify the topological nature of the quantum walk system.

The parity $\mathcal{P}$ is a unitary operator which reverses the  position as ${\bf x} \to-{\bf x}$. On the other hand, time reversal operator $\mathcal{T}$ is an anti-unitary operator which reverses the arrow of time, i.e., $t \rightarrow -t$.  A Hamiltonian $H$ is said to be parity and time reversal symmetric if it commutes with these two operators, i.e., $ \mathcal{P} H \mathcal{P}^{-1} = H,~\mathcal{T} H \mathcal{T}^{-1} = H$. The Hamiltonian $H$ is $\mathcal{PT}$-symmetric if
\begin{equation} \label{eq:PT-Symmetry}
	(\mathcal{PT}) H (\mathcal{PT})^{-1} = H.
\end{equation}
Due to the anti-linear nature of $\mathcal{PT}$ operator, even if the Hamiltonian $H$ commutes with the anti-linear operator $\mathcal{PT}$, they need not necessarily share the same set of eigenvectors. When the Hamiltonian $H$ and the anti-linear operator $\mathcal{PT}$ have the same eigenvectors, then is it called exact $\mathcal{PT}$ symmetry. In such cases the Hamiltonian possess real spectrum.

The system under consideration is quantum walk which is performed on position space with the aid of coin states. Since, the time evolution operator $U(k)$  is block diagonal in the momentum space i.e. $U(k) = \sum_{k} \tilde{U}(k) \otimes \dyad{k}$ and $U(k) = e^{-i H(k)}$, we write the condition for $\tilde{U}(k)$ in order to have $\mathcal{PT}$ symmetry as
\begin{equation}
	(\mathcal{\tilde{\mathcal{P}}\tilde{\mathcal{T}}}) \tilde{U}(k) (\mathcal{\tilde{\mathcal{P}}\tilde{\mathcal{T}}})^{-1} = \tilde{U}^{-1}(k),
\end{equation} 
where the operators $\tilde{\mathcal{P}}$, $\tilde{\mathcal{T}}$ act only on the coin Hilbert space.

In the case of non-Hermitian 1D SSQW, $\tilde{U}(k)$ given in Eq. \eqref{eq:NonUnitary-SSQW} satisfies the above mentioned conditions with the choice of $\tilde{\mathcal{P}} = \sigma_y$ and $\tilde{\mathcal{T}} = \sigma_x \mathcal{K}$ such that the combined operator becomes $\mathcal{\tilde{\mathcal{P}}\tilde{\mathcal{T}}} = i \sigma_z \mathcal{K}$ and we have
\begin{equation}
	\sigma_z \tilde{U}^*\sigma_z^{-1} = \tilde{U}^{-1}(k),
\end{equation}
where $\mathcal{K}$ is the complex conjugation operator. Therefore, the 1D SSQW is $\mathcal{PT}$-symmetric for all the values of $\delta$ (and $\gamma$). However, at the exceptional point \cite{Ozdemir2019} $\gamma_c$, the eigenstates and eigenvectors become degenerate. Beyond this point, the eigenvectors of the Hamiltonian and the $\mathcal{PT}$ operator are not the same \cite{Mochizuki2016}; hence the system no longer possesses exact-$\mathcal{PT}$-symmetry, which results in a complex spectrum, as shown in the previous section.

Next we discuss the particle hole symmetry (PHS) represented by an antiunitary operator $\Xi$, and the chiral symmetry (CS) represented by a unitary operator $\Gamma$.  For a non-Hermitian Hamiltonian, the PHS and CS symmetry conditions read~\cite{Sato2019} 
\begin{align}
	\Xi H \Xi^{-1} &= -H, \\
	\Gamma H \Gamma^{-1} &= - H^{\dagger},
\end{align}
respectively. Consequently for the time evolution operator, they read
\begin{align} 
	\Xi U(k) \Xi^{-1} &=  U(-k) \label{eq: PHS-symmetry-U}, \\
	\Gamma U(k) \Gamma^{-1} &=  U^{\dagger}(k) \label{eq: chiral-symmetry-U}.
\end{align}
We redefine the time evolution operator given in \eqref{eq:NonUnitary-SSQW} \cite{Asboth2013} by performing a unitary transformation which reads
\begin{equation} \label{eq:NonUnitary-SSQW-TS}
	\tilde{U}^{'}(k) = R(\theta_1/2) T_{\downarrow}(k) G_{\delta} R(\theta_2) T_{\uparrow}(k) G_{\delta}^{-1} R(\theta_1/2),
\end{equation}
which is related to $\tilde{U}^{^{\text{NU}}}_{_{\text{SS}}}(k)$ as $\tilde{U}^{'}(k) = R(\theta_1/2) \tilde{U}^{^{\text{NU}}}_{_{\text{SS}}}(k) R^{-1}(\theta_1/2)$. This is done to make the evolution operator symmetric in time and known as time-symmetric representation. 
The motivation behind this transformation is  to show the existence of CS and PHS in non-Hermitian 1D SSQW explicitly. We can clearly see that $\tilde{U}^{'}(k)$ satisfies Eq. \eqref{eq: PHS-symmetry-U} and Eq. \eqref{eq: chiral-symmetry-U} with the choice of $\Gamma = \sigma_x $ and $\Xi = \mathcal{K}$. Hence, with the existence of these symmetries, $\tilde{U}^{^{\text{NU}}}_{_{\text{SS}}}(k)$ belongs to a symmetry class (BDI$^{\dagger}$ \cite{Sato2019}) which supports $\mathds{Z}$ topological invariant.

\section{Results}\label{Sec:Results}
In this section, we study the behavior of the topological phases in 1D SSQW and 2D DTQW by introducing a nonzero scaling factor $\gamma$ which, essentially, makes the system non-Hermitian. In 1D SSQW, we find that the topological phases are unaffected even when the system is non-Hermitian (i.e., $\gamma \ne 0$), as far as the system possesses a real spectrum following the exact $\mathcal{PT}$-symmetry. However, the topological nature of the system vanishes  as we cross the exceptional point $\gamma_c$, which means the quantity  $W$ becomes a non-integer number which decays asymptotically to zero for $\gamma >\gamma_c$.
We observe the persistence of the Chern number $C$ in 2D DTQW as well until the scaling factor $\gamma$ reaches a critical value. However, unlike the 1D case, we cannot associate exact $\mathcal{PT}$-symmetry breaking with the point where the topological phase transition happens due to the absence of the $\mathcal{PT}$-symmetry in 2D DTQW. Since the $\mathcal{PT}$-symmetry is absent in 2D DTQW even in the unitary region, we can not associate the persistence of the topological phase with this particular symmetry. Furthermore, we observe a loss-induced topological phase transition in 2D DTQW.

\subsection{Topological phases in 1D non-unitary quantum walk}
We start our analysis with non-unitary 1D SSQW, with the associated non-Hermitian Hamiltonian $H_{_{\text{NU}}}(\theta_1, \theta_2, \gamma)$ being given in \eqref{eq:Hamil-SSQW}. Since the Hamiltonian is traceless for all values of $\gamma$, the corresponding eigenvalues will always be of the form of $\pm E(k)$. For each momentum $k$, we compute the energy eigenstates $\ket{\psi_\pm(k)}$ corresponding to energies $\pm E(k)$ and, we call the set $\{\ket{\psi_-(k)}\}$ and $\{\ket{\psi_+(k)}\}$ as the lower and upper energy bands, respectively. Using the expression for the winding number $W$ from \eqref{eq: Winding-Number}, we calculate the winding numbers for the lower and upper bands and we name them as $W_-$ and $W_+$, respectively. 

\begin{figure*}
	\centering
	\subfigure[]{
		\includegraphics[width=3.5cm]{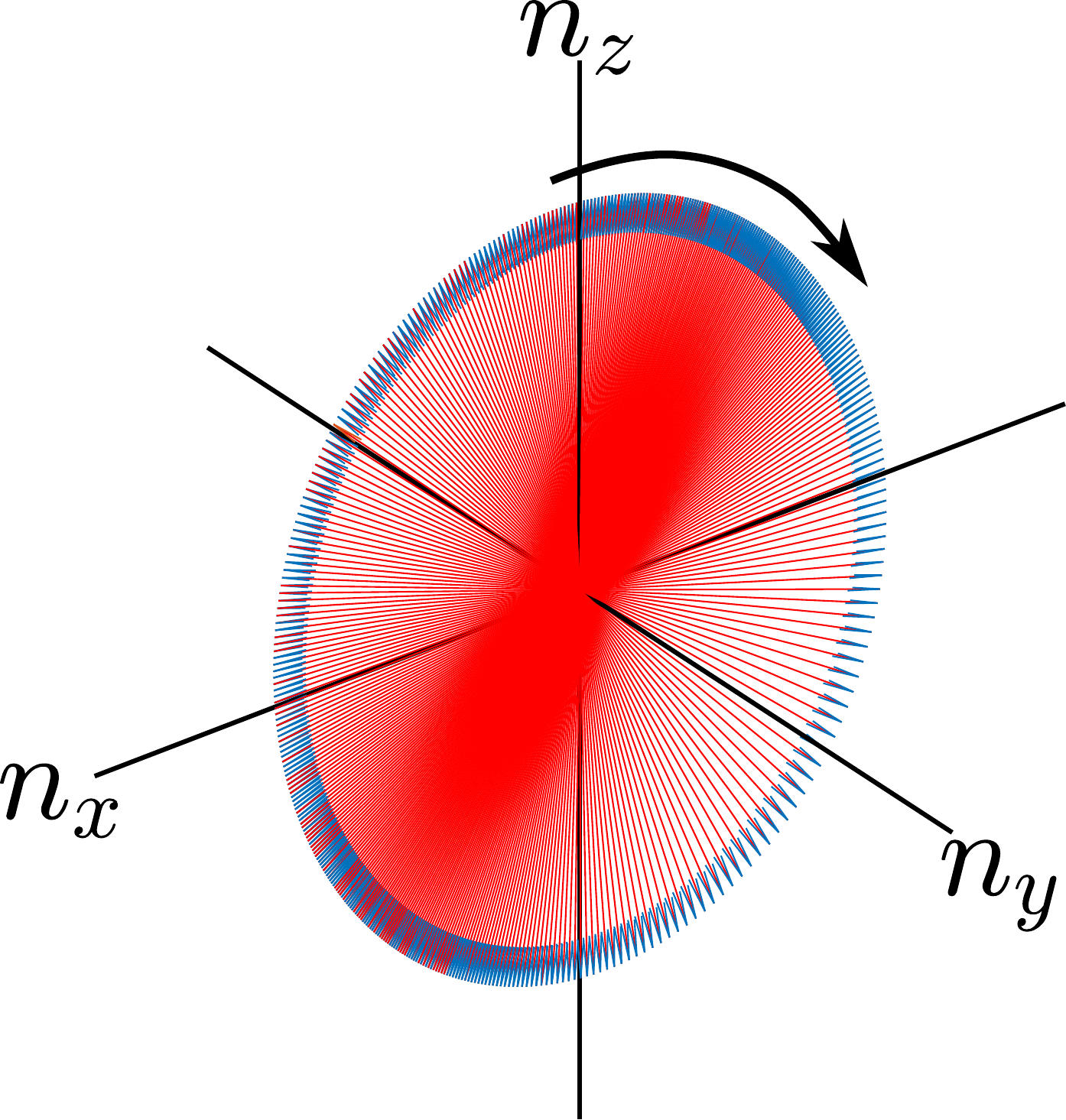}
		\label{fig:NVEC1}}
	\subfigure[]{
		\includegraphics[width=3.5cm]{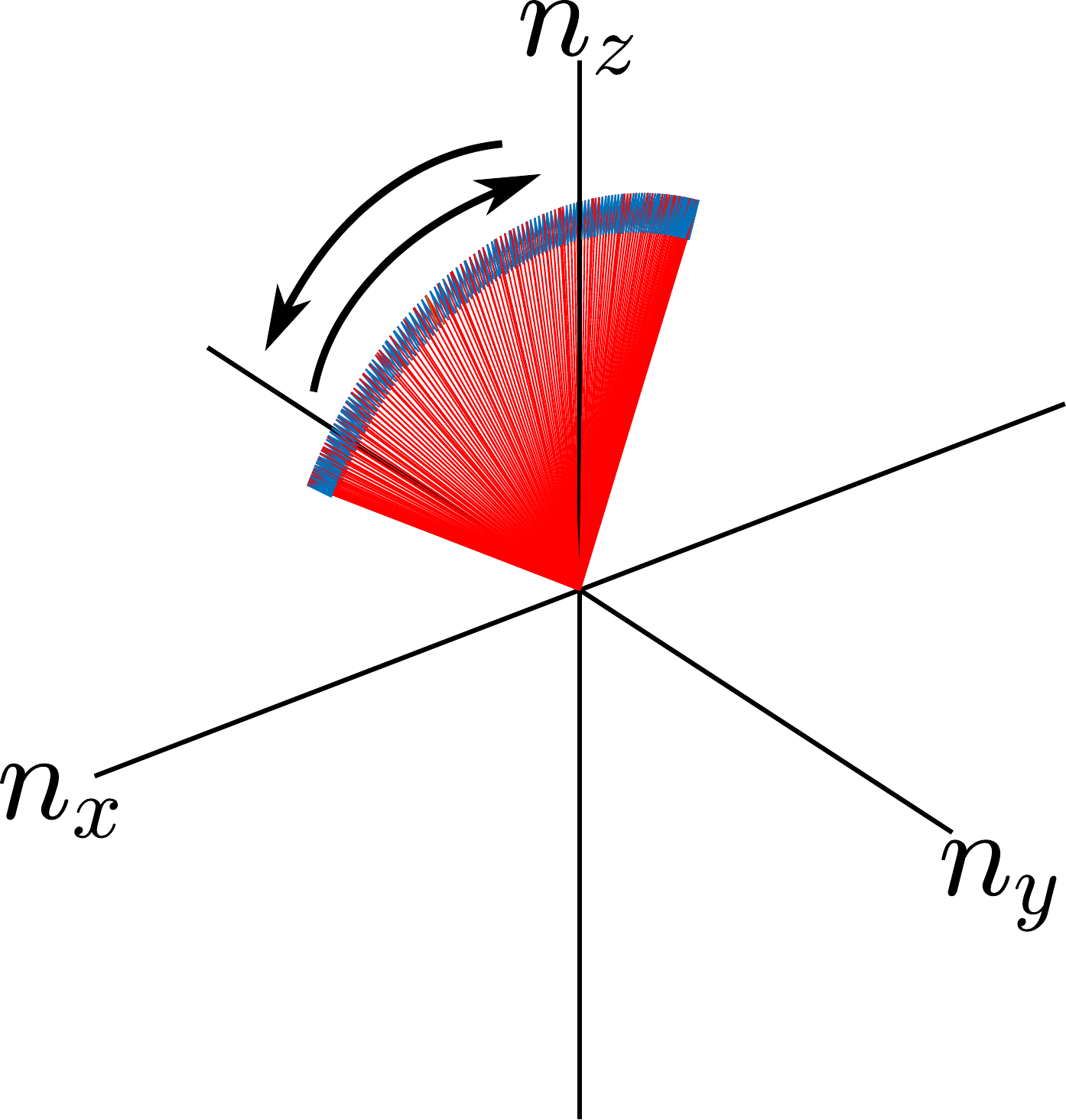}
		\label{fig:NVEC2}} 
	\subfigure[]{
		\includegraphics[width=3.5cm]{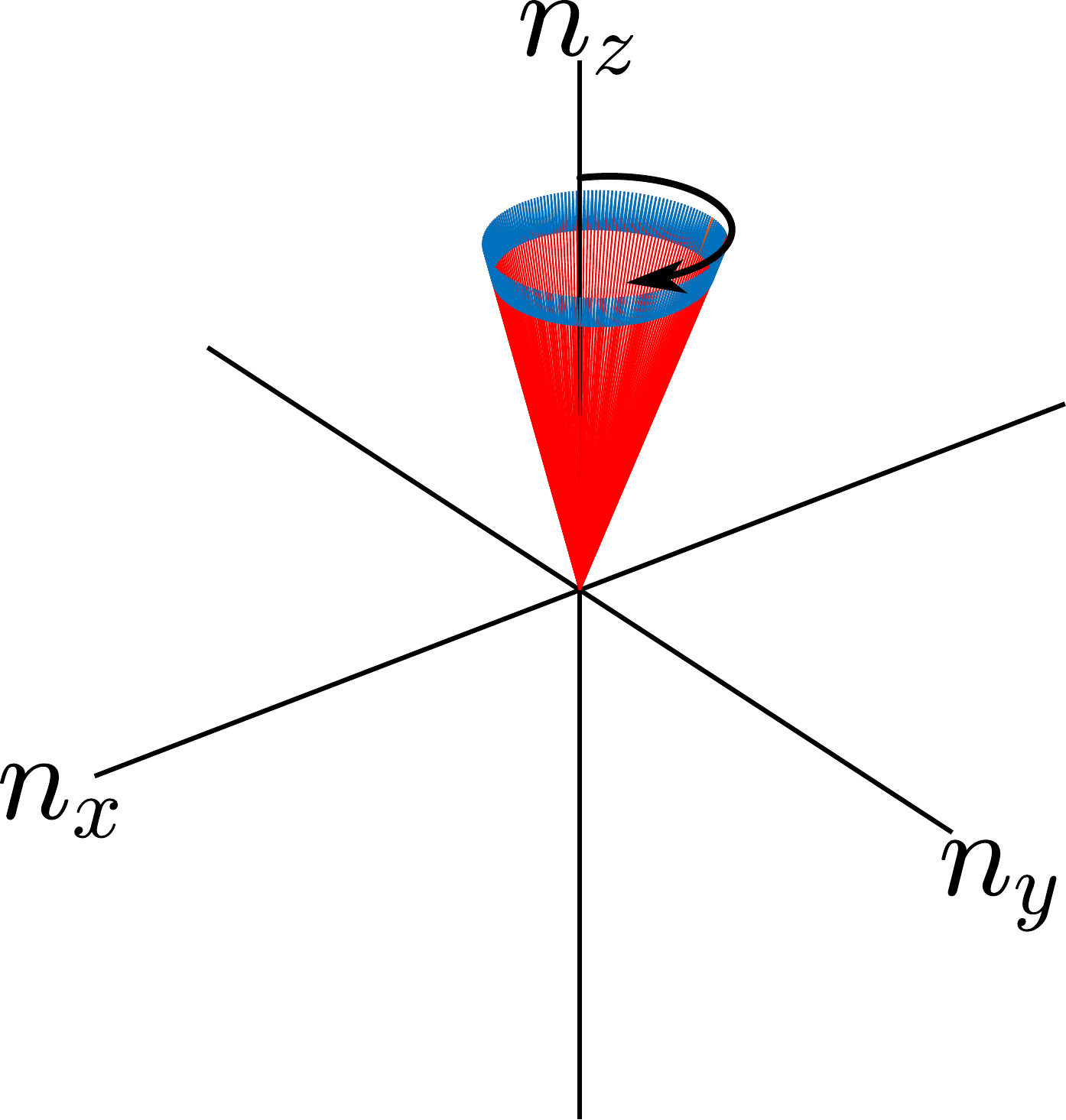}
		\label{fig:NVEC3}}
	\subfigure[]{
		\includegraphics[width=3.5cm]{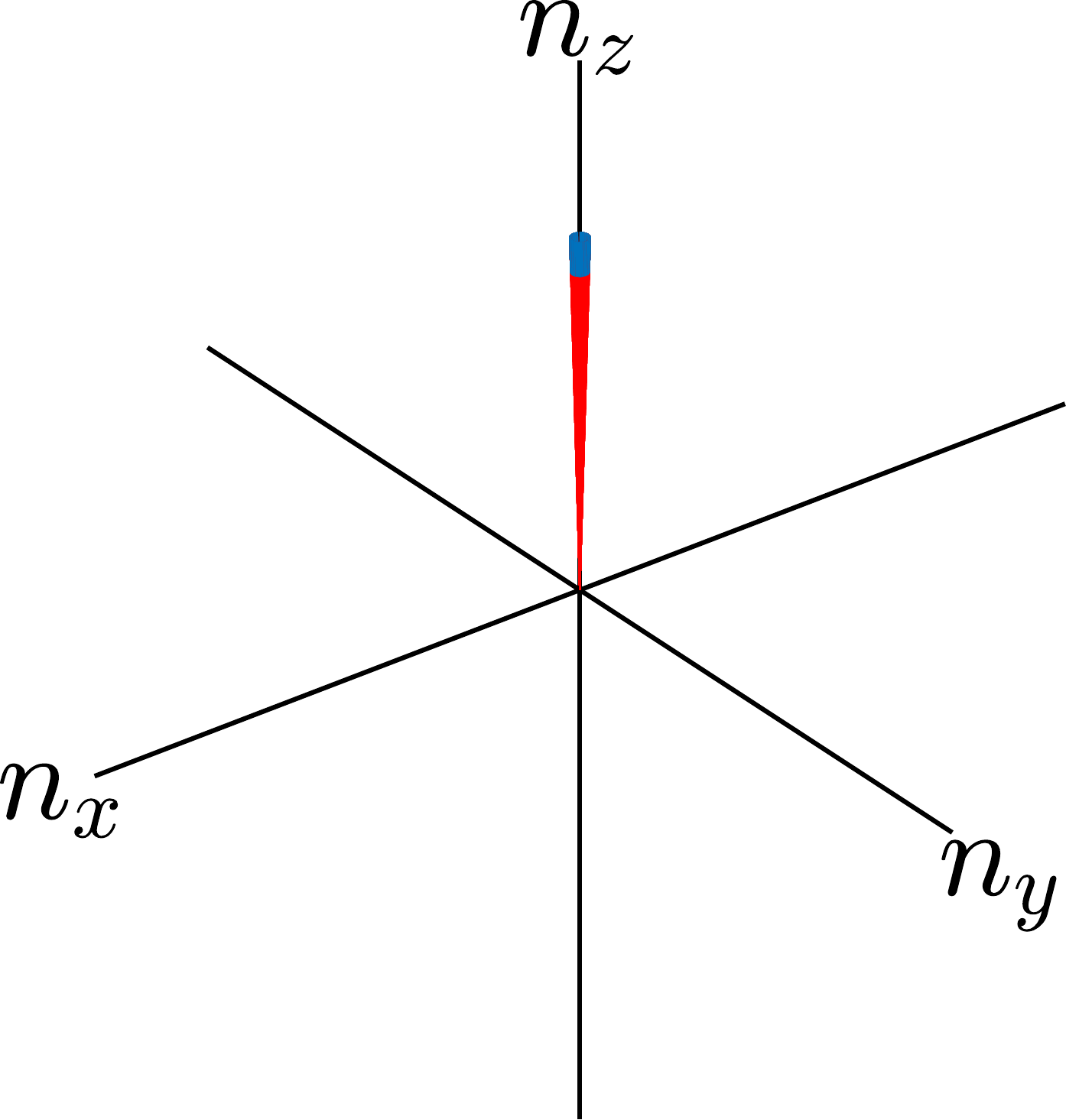}
		\label{fig:NVEC4}}
	\caption{(Color online) Winding of the Bloch vector around the origin with the lattice size, $N = 201~ \subref{fig:NVEC1}~\theta_1 = -3\pi/8,~\theta_2 = \pi/8,~\gamma=0.25~\subref{fig:NVEC2}~\theta_1 = -3\pi/8,~\theta_2 = 5\pi/8,~\gamma=0.25~\subref{fig:NVEC3}~\theta_1 = -3\pi/8,~\theta_2 = \pi/8, \gamma = 1.8~\subref{fig:NVEC4}~\theta_1 = -3\pi/8,~\theta_2 = \pi/8, \gamma = 3.0$.}
	\label{fig:NVEC}
\end{figure*}

Since the eigenstates and eigenvalues depend on $\gamma$, $\theta_1$ and $\theta_2$,  the winding numbers are also expected to depend upon these parameters. In Fig.\,\ref{fig:SSQWL}, we plot the  winding number for the lower band $W_-$ as a function of $\gamma$ and $\theta_2$ for different values of $\theta_1$. In all figures, we notice that for $\gamma=0$, the winding number can take two different values, zero and one, depending on the choice of $\theta_1$ and $\theta_2$. Focusing on the case of $W_-=1$ for a vanishing $\gamma$, we observe that for a given $(\theta_1,\theta_2)$ if we increase the value of $\gamma$, the winding number remains unaffected until we approach the critical value of $\gamma$, i.e., $\gamma_c$ \eqref{Eq:Delta-c}. Crossing the $\gamma_c$ causes a phase transition and the value of $W$ starts decreasing and approaches zero asymptotically.  Whereas, if initially the winding number $W_-= 0$, it remains zero until we approach $\gamma_c$, and then it starts to increase momentarily approaching some maximum value and then deteriorates to zero asymptotically.

By definition, the winding number is an integer quantity. In other words, the geometric phase acquired by the eigenstates of the Hamiltonian in the $k$-space is quantized and is a multiple of $\pi$, which is possible only when all the states in an energy band lie in a plane on the Bloch sphere. The winding number must always be an integer for all the Hermitian Hamiltonians. However, beyond the exceptional points, $W$ acquires non-integer values, hence it does not qualify as the winding number. This non-integer value of $W$ can be explained by observing the behaviour of the eigenstates of the non-Hermitian Hamiltonian. In Fig.\,\ref{fig:NVEC}, we plot the Bloch vectors corresponding to the eigenstates $\ket{\psi_-(k)}$ of the Hamiltonian $H_{_{\text{NU}}}(\theta_1,\theta_2,\gamma)$ on the Bloch sphere. In Fig.~\ref{fig:NVEC1}, the Bloch Vector moves in the clockwise direction and comes back to the same point, winding around the origin once resulting in $W=1$. However, in Fig. \ref{fig:NVEC2}, it first goes clockwise and reverses its direction, and; therefore, $W=0$. Figs.~\ref{fig:NVEC1} and \ref{fig:NVEC2} are for $\gamma \le \gamma_c$ whereas Figs.~\ref{fig:NVEC3} and \ref{fig:NVEC4} are for $\gamma > \gamma_c$. The animation of Bloch vectors can be found in the Supplementary Material online. We can clearly see that in the exact $\mathcal{PT}$-symmetric region, the eigenstates lie in a plane and results in an integer value of $W$, whereas in the exact $\mathcal{PT}$-symmetry broken region the eigenvectors trace a path which lies outside the plane. Hence geometric phase is not a multiple of $\pi$ resulting in a non-integer value of $W$. 

In summary, we have shown that the topological phase in 1D SSQW remains invariant as long as the energy eigenvalues are real, even though the Hamiltonian is not Hermitian, i.e., the topological order persists as long as the Hamiltonian respects exact $\mathcal{PT}$ symmetry. Next, we extend our study to the case of 2D DTQW.
\begin{figure*}
	\centering
	\subfigure[]{
		\includegraphics[width=5cm]{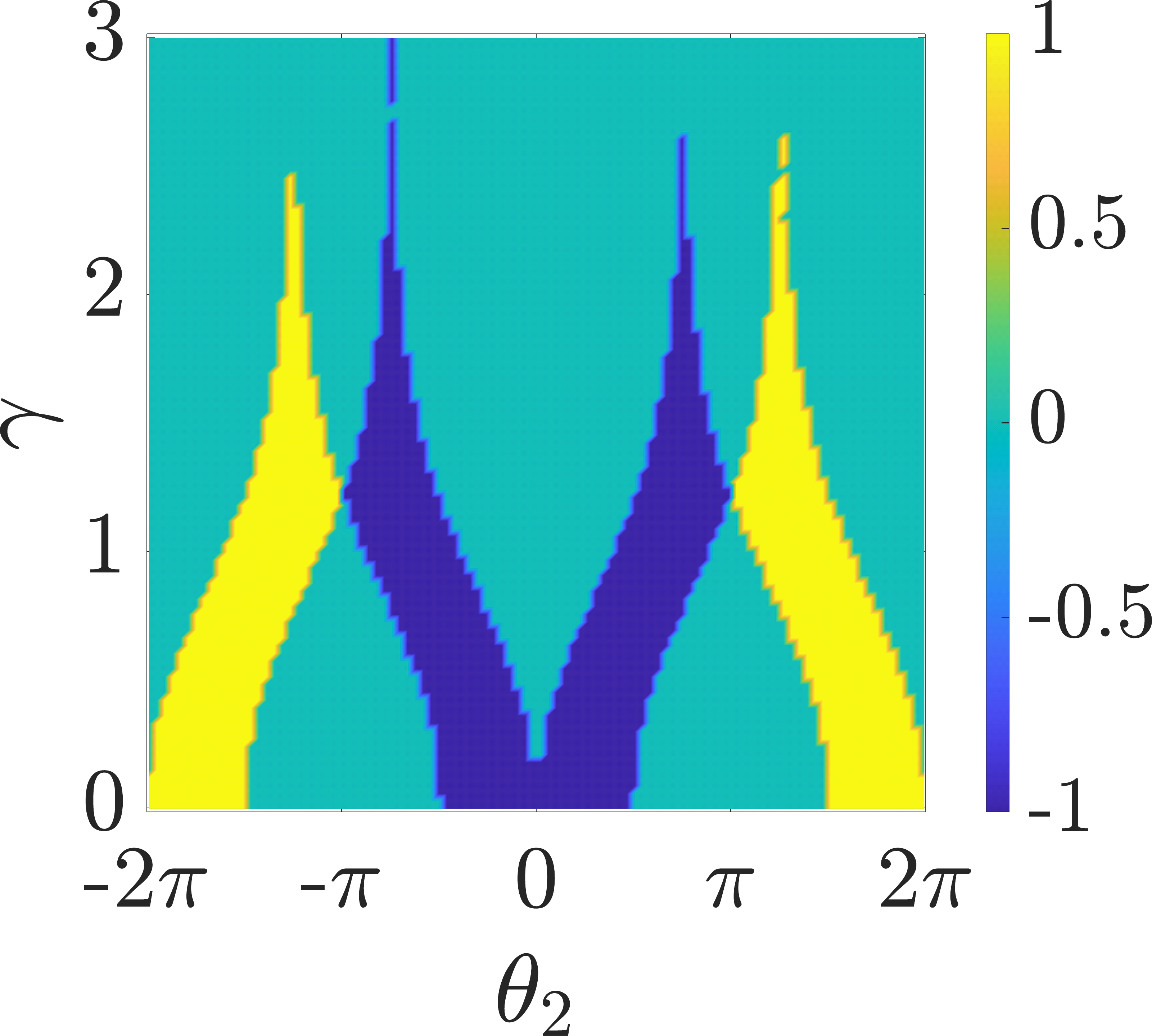}
		\label{fig:DTQWpiby4a}}
	\subfigure[]{
		\includegraphics[width= 5cm]{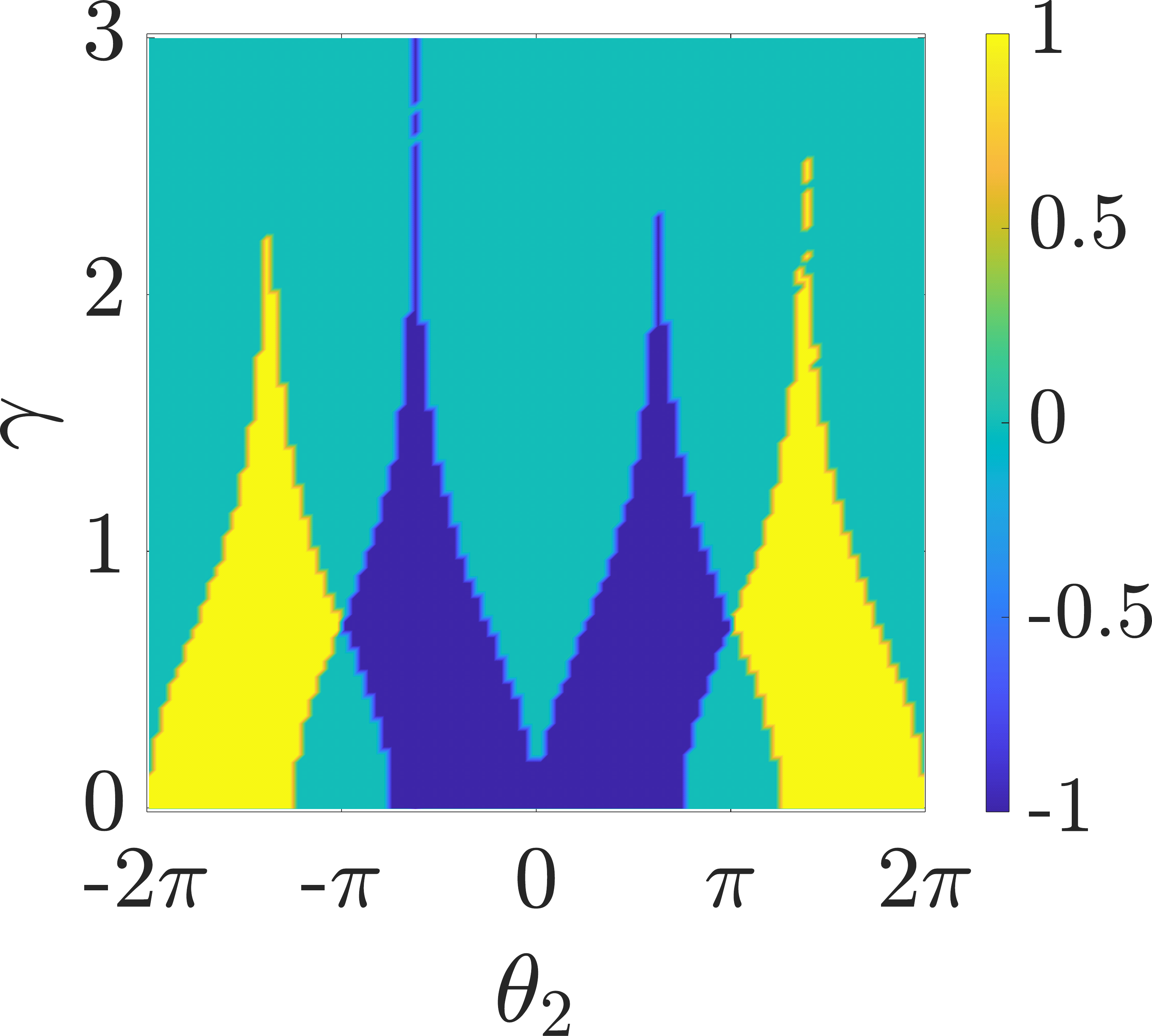}
		\label{fig:DTQW3piby8a}}
	\subfigure[]{
		\includegraphics[width= 5cm]{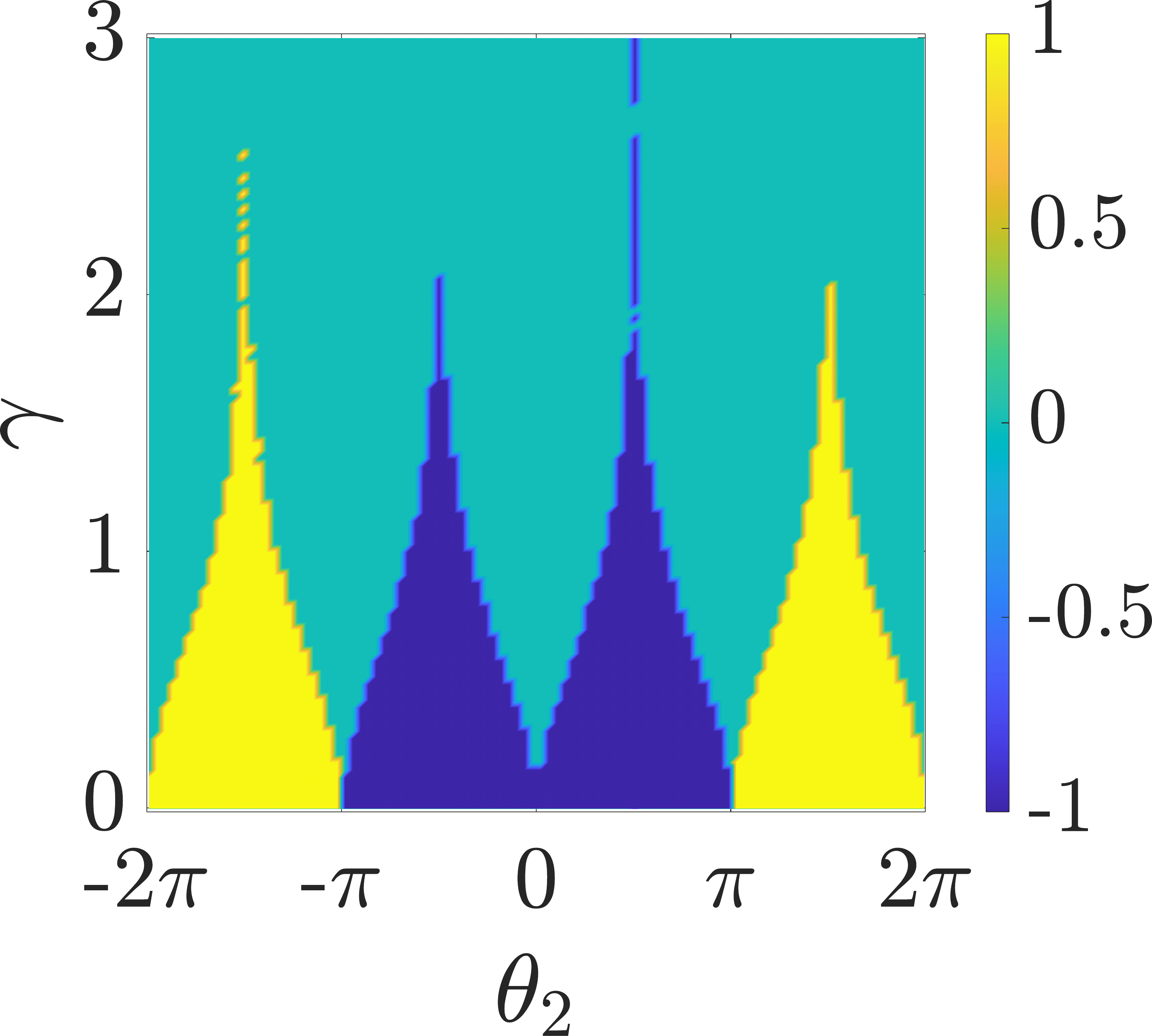}
		\label{fig:DTQW3piby2a}}
	\subfigure[]{
		\includegraphics[width=5cm]{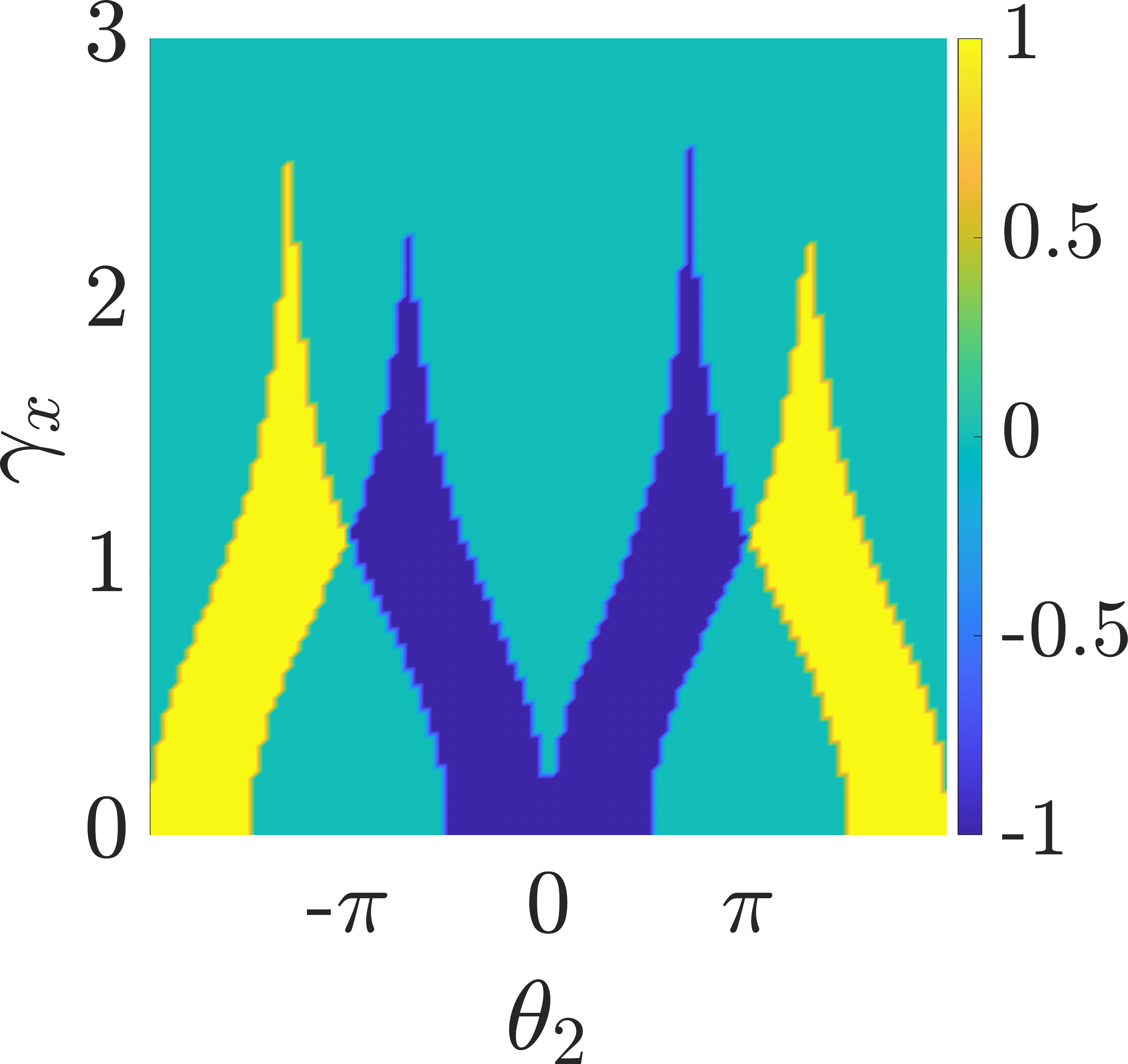}
		\label{fig:DTQWpiby4b}}
	\subfigure[]{
		\includegraphics[width= 5cm]{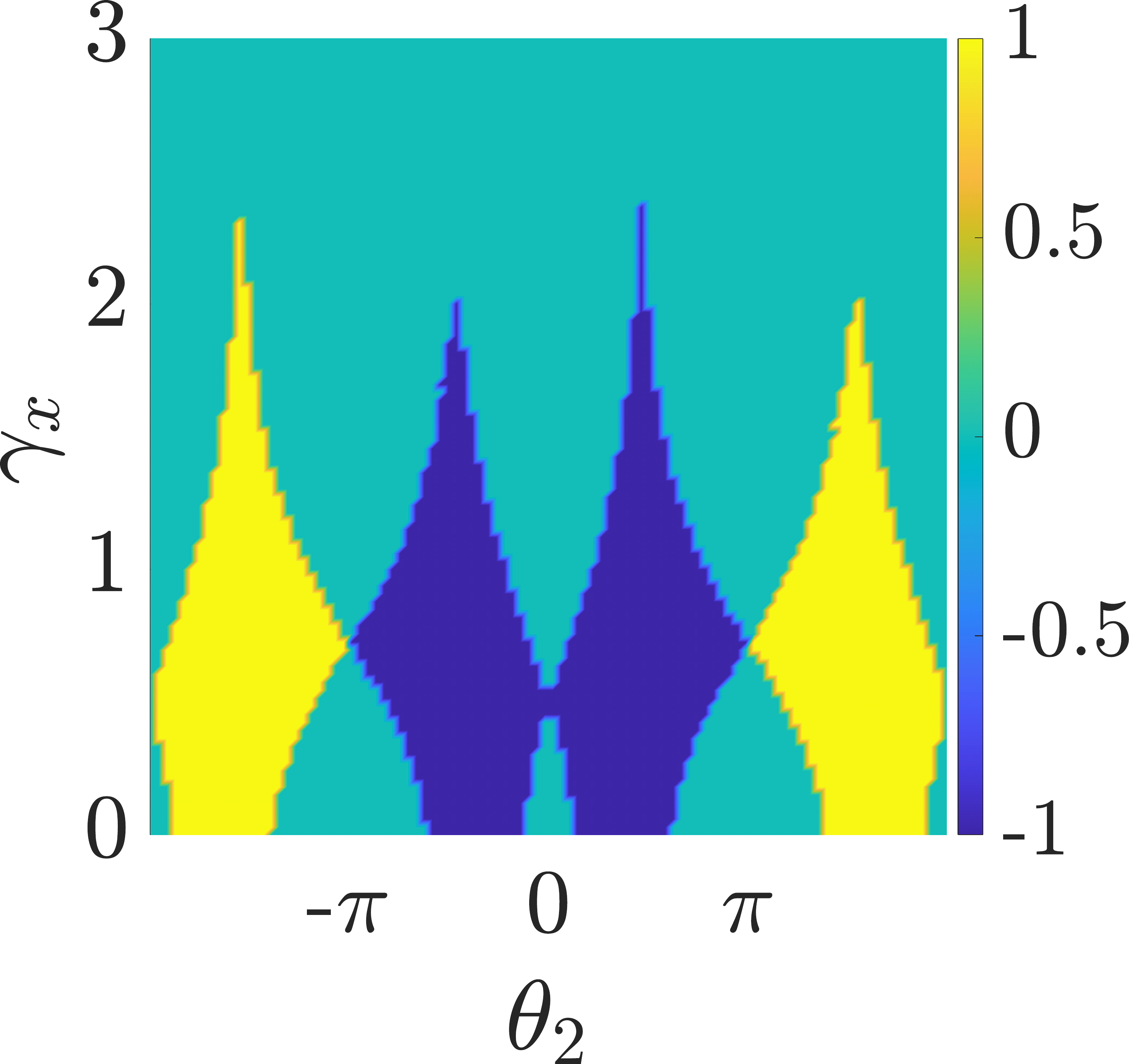}
		\label{fig:DTQW3piby8b}}
	\subfigure[]{
		\includegraphics[width= 5cm]{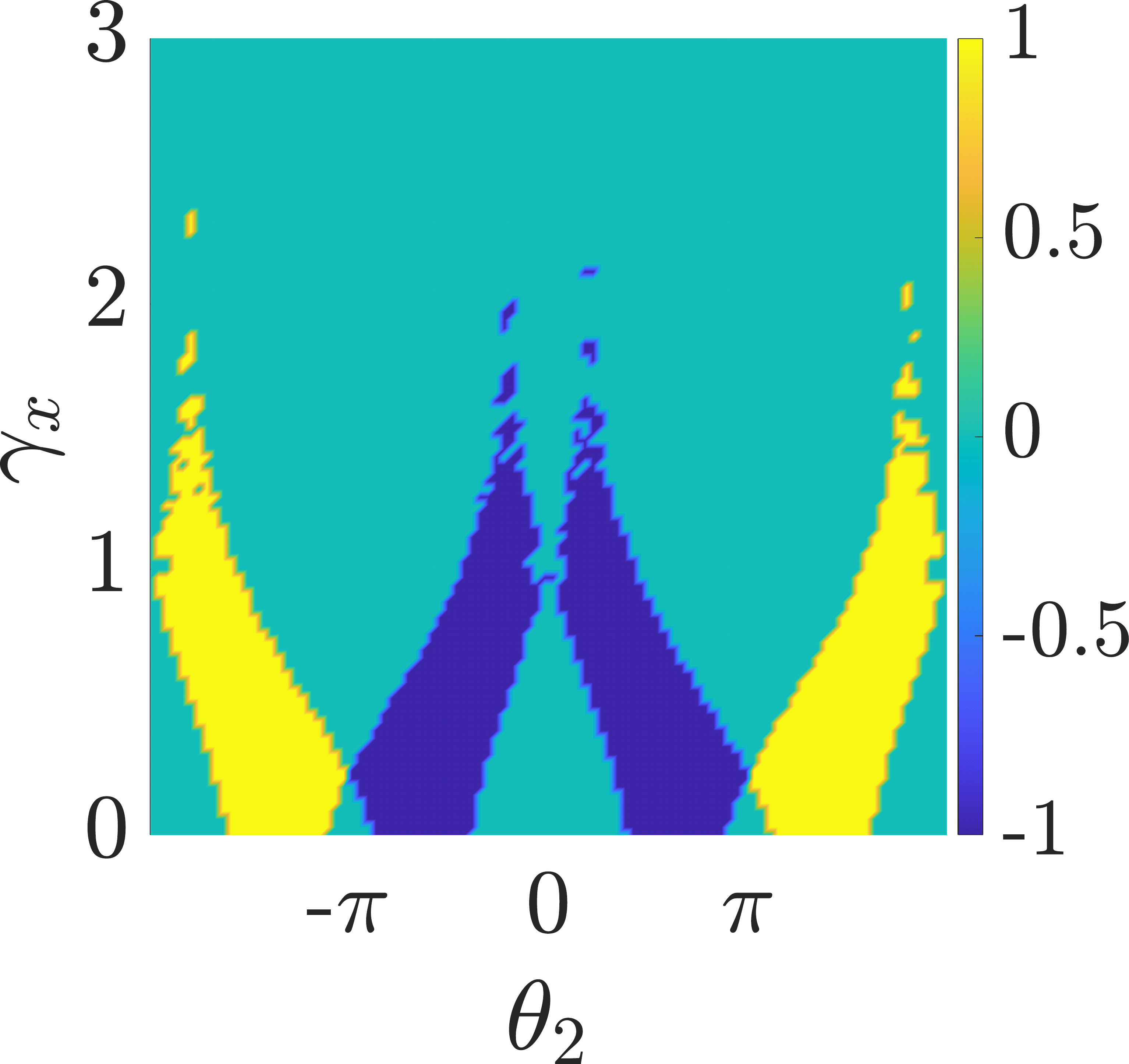}
		\label{fig:DTQW3piby2b}}
	\caption{(Color online) Effect of $\gamma_x$ on Chern number is plotted with varying $\theta_2$ for $\gamma_y = 0$ \subref{fig:DTQWpiby4a} $\theta_1 = \pi/4$ \subref{fig:DTQW3piby8a} $\theta_1 = 3\pi/8$ \subref{fig:DTQW3piby2a} $\theta_1 = 3\pi/2$. In the bottom row, \subref{fig:DTQWpiby4b} $\gamma_y = 0.1$, \subref{fig:DTQW3piby8b} $\gamma_y = 0.5$, \subref{fig:DTQW3piby2b} $\gamma_y = 1.0$, respectively. The lattice size is taken to be 201 $\times$ 201.}
	\label{fig:2DDTQWG}
\end{figure*}

\subsection{Topological phases in 2D non-unitary quantum walk}
Since 2D DTQW can be decomposed as a product of two 1D SSQW, we can easily extend 2D DTQW to non-unitary limits by introducing the scaling operator  $G$ along the $x$- as well as the $y$-axis. The time evolution operator can be written as
\begin{equation} \label{eq:Nonunitary-2DDTQW}
	U^{^{\text{NU}}}_{_{2D}}(\theta_1, \theta_2,\gamma_x, \gamma_y) = G_{\gamma_y} T_y R(\theta_1) G_{\gamma_y}^{-1} T_y R(\theta_2) G_{\gamma_x} T_x R(\theta_1) G_{\gamma_x}^{-1}T_x.
\end{equation}
The corresponding non-Hermitian Hamiltonian of this system reads
\begin{align}
	H^{^{\text{NU}}}_{_{2D}}(\theta_1, \theta_2,\gamma_x, \gamma_y) = \bigoplus_{k_x,k_y} E(k_x,k_y,\gamma_x, \gamma_y) {\bf n}(k_x,k_y,\gamma_x, \gamma_y)\cdot {\bf \sigma},
\end{align}
where
\begin{align}
	&\cos E(k_x, k_y,\gamma_x, \gamma_y) \nonumber \\
	=  & \cos \theta_1 \cos (\theta_2/2) \cos (k_x + k_y - i \gamma_x + i \gamma_y) \cos (k_x + k_y + i \gamma_x - i \gamma_y) \nonumber \\
	&- \cos (\theta_2/2) \sin(k_x + k_y - i \gamma_x + i \gamma_y)\sin(k_x + k_y +i \gamma_x - i \gamma_y) \nonumber \\
	&- \sin \theta_1\sin (\theta_2/2) \cos(k_x - k_y - i \gamma_x - i \gamma_y) \cos(k_x + k_y + i \gamma_x - i \gamma_y),
\end{align}
and
\begin{equation}
	\hat{\vb{n}}(k_x,k_y,\gamma_x, \gamma_y) = \dfrac{n_x(k_x,k_y,\gamma_x, \gamma_y) \hat{\vb{i}} + n_y(k_x,k_y,\gamma_x, \gamma_y) \hat{\vb{j}} + n_z(k_x,k_y,\gamma_x, \gamma_y) \hat{\vb{k}}}{\sin E(k_x, k_y,\gamma_x, \gamma_y)},
\end{equation}
with 
\begin{align}
	n_x(k_x, k_y,\gamma_x, \gamma_y) =& - \sin \theta_1\cos(\theta_2/2) \cos(k_x + k_y -i \gamma_x + i \gamma_y) \sin (k_x - k_y +i \gamma_x + i \gamma_y) \nonumber \\
	&- \cos \theta_1 \sin(\theta_2/2) \cos (k_x - k_y - i \gamma_x - i \gamma_y) \sin (k_x - k_y + i \gamma_x + i \gamma_y) \nonumber \\
	&- \sin(\theta_2/2)  \sin (k_x - k_y - i \gamma_x - i \gamma_y) \cos (k_x - k_y + i \gamma_x + i \gamma_y), \nonumber \\ 
	n_y(k_x, k_y,\gamma_x, \gamma_y)=& \sin \theta_1\cos(\theta_2/2) \cos(k_x + k_y - i \gamma_x + i \gamma_y) \cos (k_x - k_y + i \gamma_x + i \gamma_y) \nonumber \\
	&+ \cos \theta_1 \sin(\theta_2/2) \cos (k_x - k_y - i \gamma_x - i \gamma_y) \cos (k_x - k_y + i \gamma_x + i \gamma_y ) \nonumber \\
	&- \sin(\theta_2/2)  \sin (k_x - k_y - i \gamma_x - i \gamma_y) \sin (k_x - k_y + i \gamma_x + i \gamma_y), \nonumber \\
	n_z(k_x, k_y,\gamma_x, \gamma_y)=& -\cos \theta_1 \cos(\theta_2/2) \cos (k_x + k_y - i \gamma_x + i \gamma_y )\sin (k_x + k_y + i \gamma_x - i \gamma_y) \nonumber \\
	&-\cos(\theta_2/2) \sin (k_x + k_y - i \gamma_x + i \gamma_y) \cos (k_x + k_y + i \gamma_x - i \gamma_y) \nonumber \\
	&+ \sin \theta_1 \sin(\theta_2/2) \cos(k_x - k_y-i \gamma_x - i \gamma_y) \sin(k_x + k_y+ i \gamma_x - i \gamma_y) \nonumber.
\end{align}

The 2D DTQW is different from the 1D SSQW as the former case does not support $\mathcal{PT}$-symmetry even in the unitary region. The energy eigenvalues become complex even for very small values of the scaling factor. 
If we take $\gamma_x << 1$ and $\gamma_y = 0$, the expression for the energy reads
\begin{equation} \label{eq: No PT-Symmetry}
	\cos E(\gamma_x) = \cos E(\gamma_x = 0) + i \gamma_x \sin \theta_1 \sin (\theta_2/2) \sin (2 k_y),
\end{equation}
which makes the quasi-energy complex for infinitesimal scaling parameter $\gamma_x$. 

For 2D DTQW we will have $\vb{k} = (k_x,k_y)$ and the time evolution operator in Eq. \eqref{Eq:Qwalk2D} in momentum space must satisfy $	\Xi U(\vb{k}) \Xi^{-1} = U(-\vb{k})$ in order to possess PHS \cite{Kitagawa2010, Mochizuki2016}, which is satisfied by choosing $\Xi = \mathcal{K}$ for all the values of scaling factor $G_{\gamma_x}$ and $G_{\gamma_y}$. The existence of PHS ensures that 2D DTQW realizes topological phases with $\mathds{Z}$ topological invariant \cite{Schnyder2008,Kitaev2009}.

Similar to the case of 1D SSQW, in 2D quantum walks also the energy eigenvalues appear in pairs $\pm E(k_x, k_y,\gamma_x, \gamma_y)$ resulting in two energy bands. Introducing loss and gain (scaling factor $\gamma$) in $x$ and $y$-direction results in complex pairs of energy eigenvalues. We can choose the lower energy state by looking at the sign of the real part of the energy eigenstate and calculate the Chern number. 

We use \eqref{eq:Chern-Number} to calculate the Chern number for the lower energy band and plot it against  $\gamma_x$ and $\theta_2$ for some fixed values of $\theta_1$ and $\gamma_y$ (Fig.\,\ref{fig:2DDTQWG}). Despite the absence of a real spectrum, we see the persistence of the topological phase as we turn on the scaling . In other words, the system remains in the same topological phase as we introduce loss and gain factors. 
In 2D DTQW  we observe another interesting feature, namely, for some particular values of $\theta_1$ and $\theta_2$, the Chern number can change abruptly from one integer value to another as $\gamma_x$ increases, resulting in a topological phase transition. This is a loss-induced topological phase transition. Furthermore, unlike the 1D SSQW, the Chern number in 2D DTQW changes abruptly and for sufficiently large values of  $\gamma_x$ and $\gamma_y$ the Chern number for all the parameters becomes zero.

\subsection{Bulk-boundary correspondence}
\begin{figure*}
	\centering
	\subfigure[]{
		\centering
		\includegraphics[width=7cm]{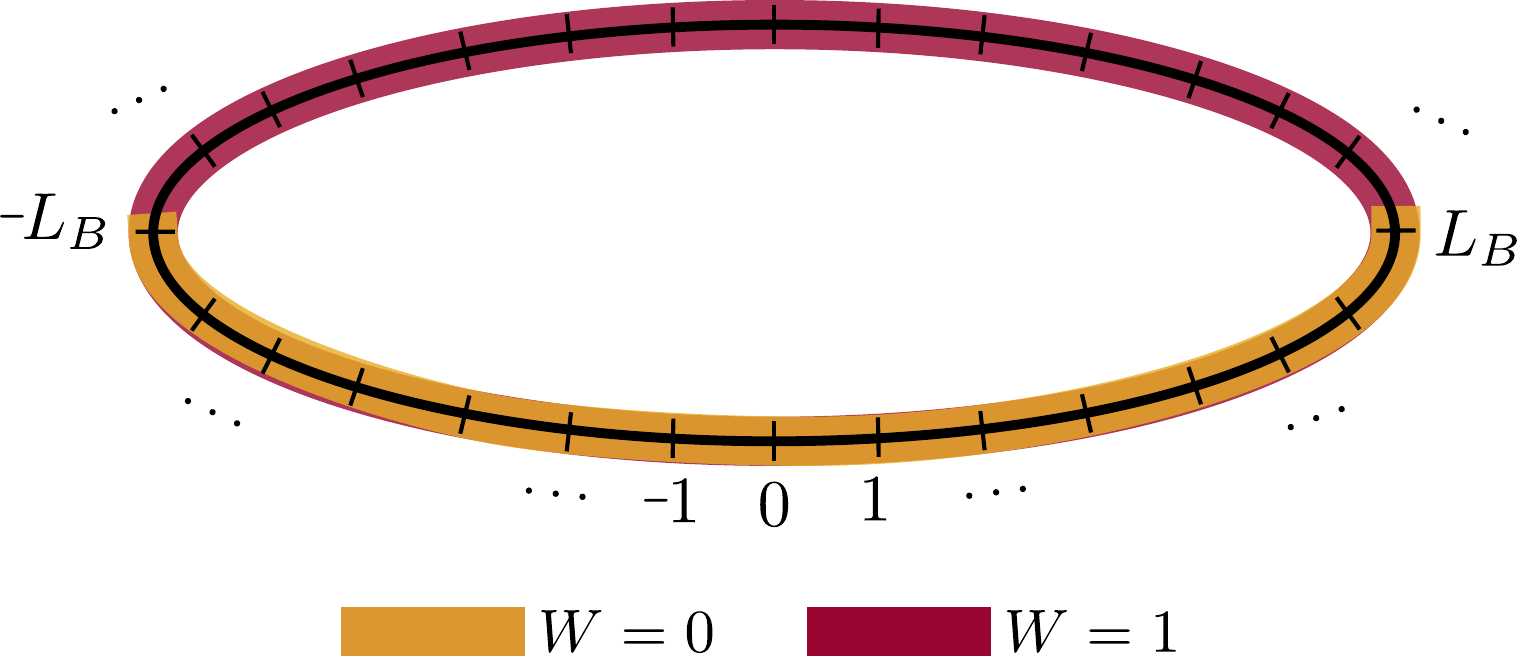}
		\label{fig:BulkEdge1D}}
	\hspace{2cm}
	\subfigure[]{
		\centering
		\includegraphics[width=5cm]{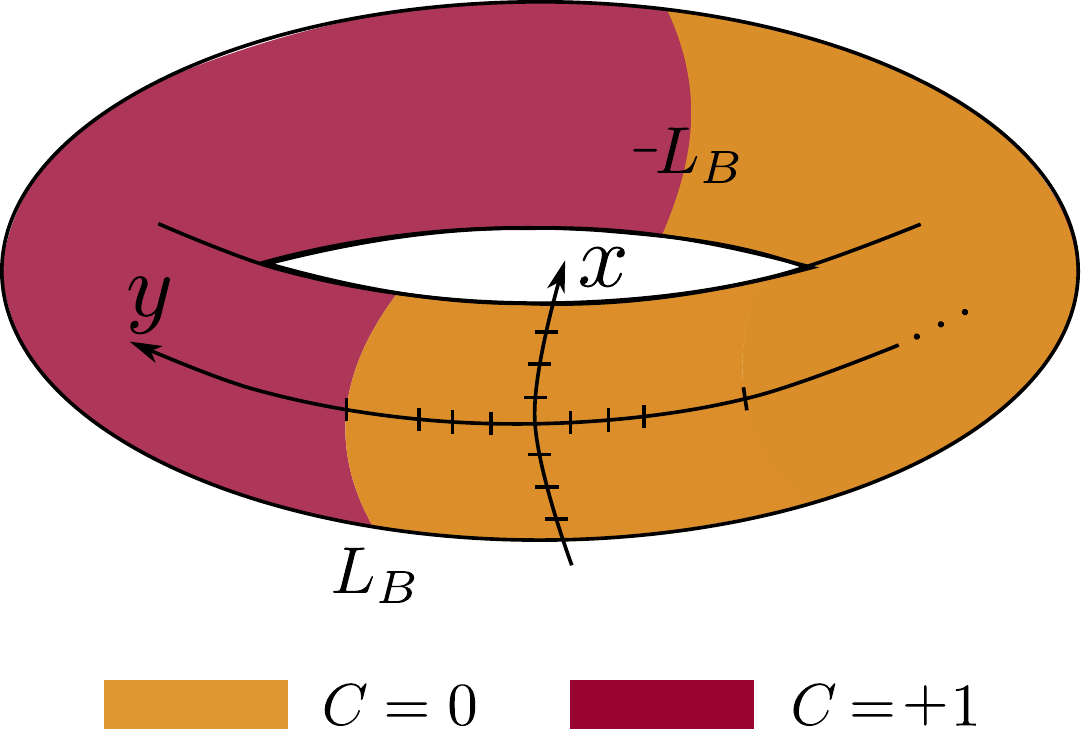}
		\label{fig:BulkEdge2D}}
	\caption{(Color online) The bulk boundary correspondence is studied by dividing the lattice in two parts which are characterized by distinct topological phases locally. \subref{fig:BulkEdge1D} 1D lattice is divided in two equal parts with two boundaries at $\pm L_B$ with $L_B = 50$ and system size $N = 201$. \subref{fig:BulkEdge2D} A two dimensional periodic lattice is divided into two equal parts where the partition is made in the $y$-direction while retaining the periodicity in the $x$-direction. The boundary on the $y$-axis is chosen at $\pm L_B$ with a lattice size 201 $\times$ 201.}
	\label{fig:BulkEdge}
\end{figure*}

In the case of infinite lattice or with periodic boundary condition, we characterize our system with topological invariants such as Winding number and Chern number, however, when we have finite lattice with open boundary conditions, we observe topologically protected states on the boundary \cite{Asboth2016, Kane2013}. In the bulk of topological insulators, the system behaves like an ordinary insulator but on the edges, we find conducting edges states. This is referred to as bulk-boundary correspondence. In this section, we study the edge states in the 1D SSQW and the 2D DTQW systems to ensure the persistence of the topological states and hence topological order.

The 1D SSQW is generally performed on an infinite lattice or a closed chain. In order to create a boundary in this system, we still consider the quantum walk on a closed chain, but divide the lattice into two regions with different  rotation angles $(\theta_1, \theta_2)$, thus making the lattice inhomogeneous. The parameters for the two parts are chosen such that the two parts locally have different topological phases, as shown in Fig.~\ref{fig:BulkEdge1D}. At the boundary of these two phases, we should see edge states which establish the topological nature of 1D SSQW.

\begin{figure*}
	\centering
	\subfigure[]{
		\includegraphics[width=4.1cm]{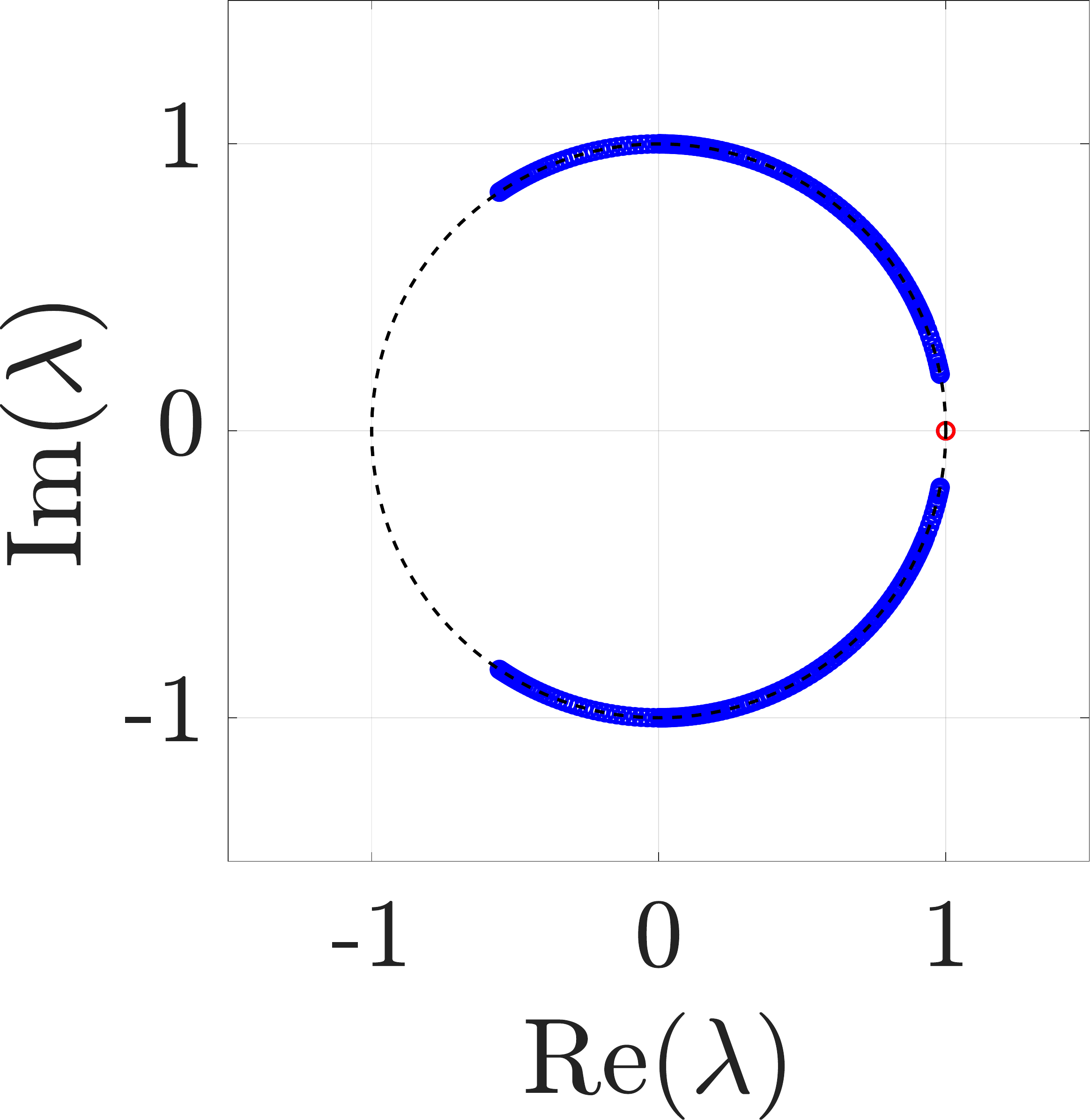}
		\label{fig:1D1a}}
	\subfigure[]{
		\includegraphics[width=4.1cm]{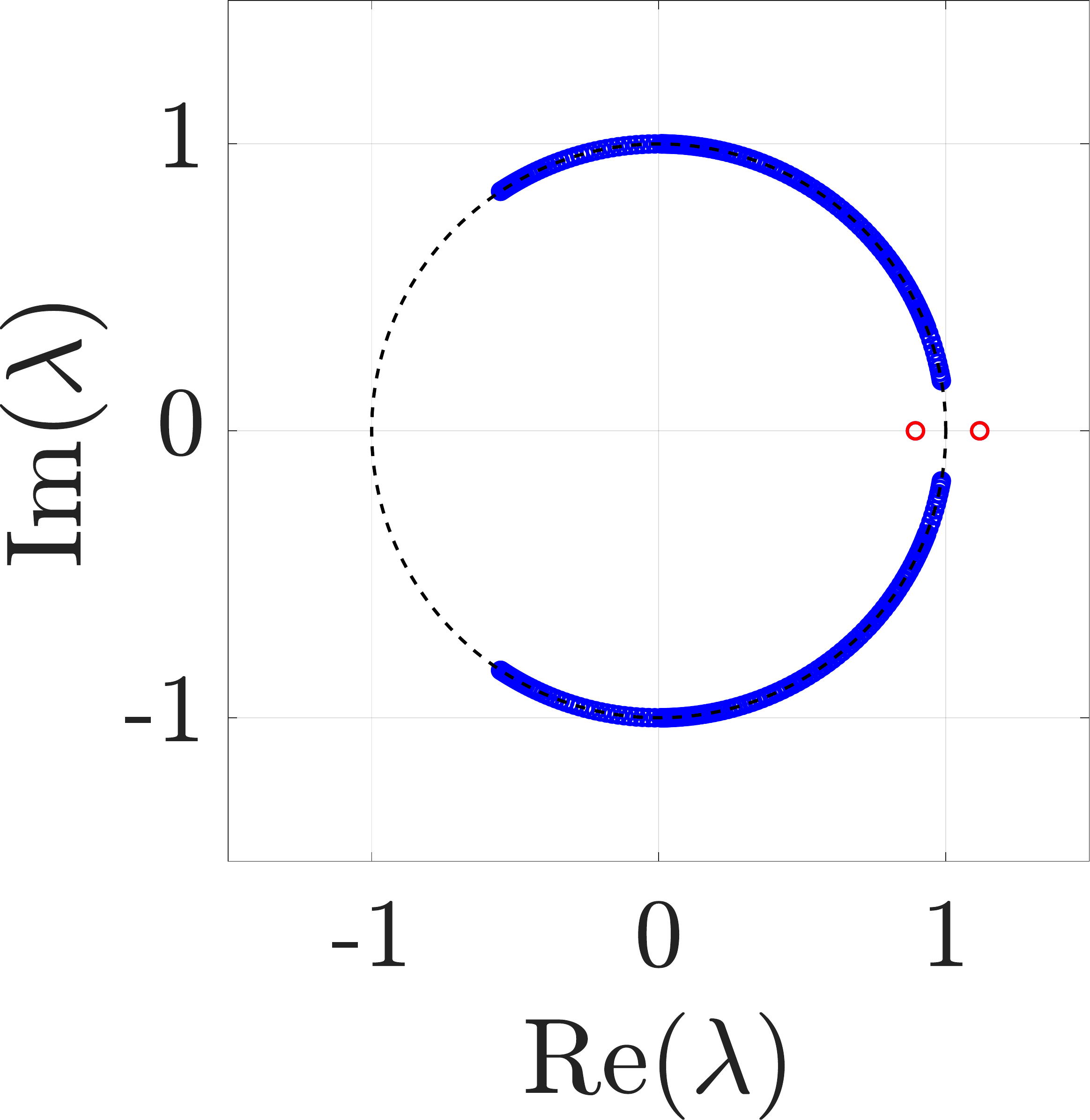}
		\label{fig:1D2a}}
	\subfigure[]{
		\includegraphics[width=4.1cm]{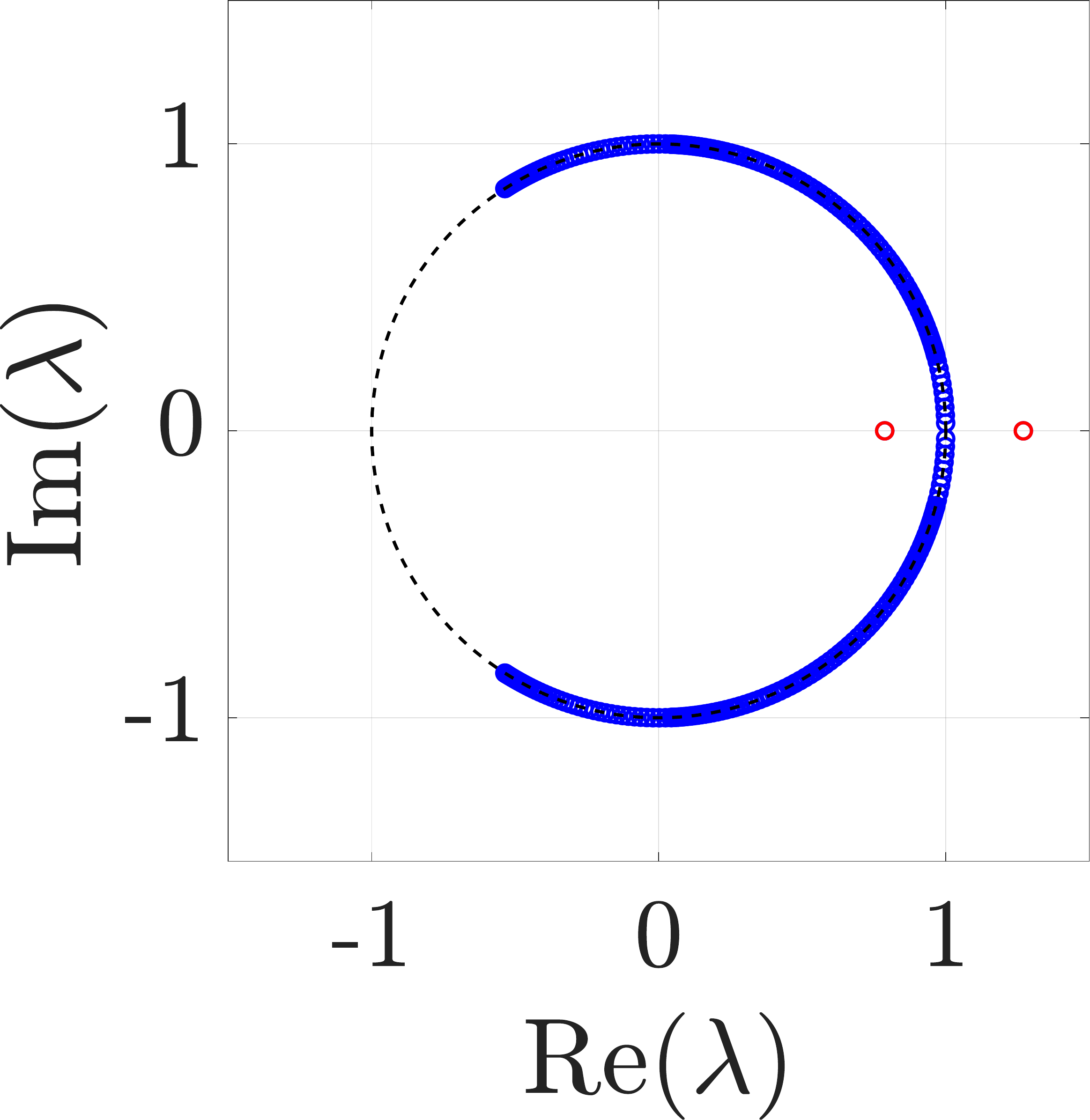}
		\label{fig:1D3a}}
	\subfigure[]{
		\includegraphics[width=4.1cm]{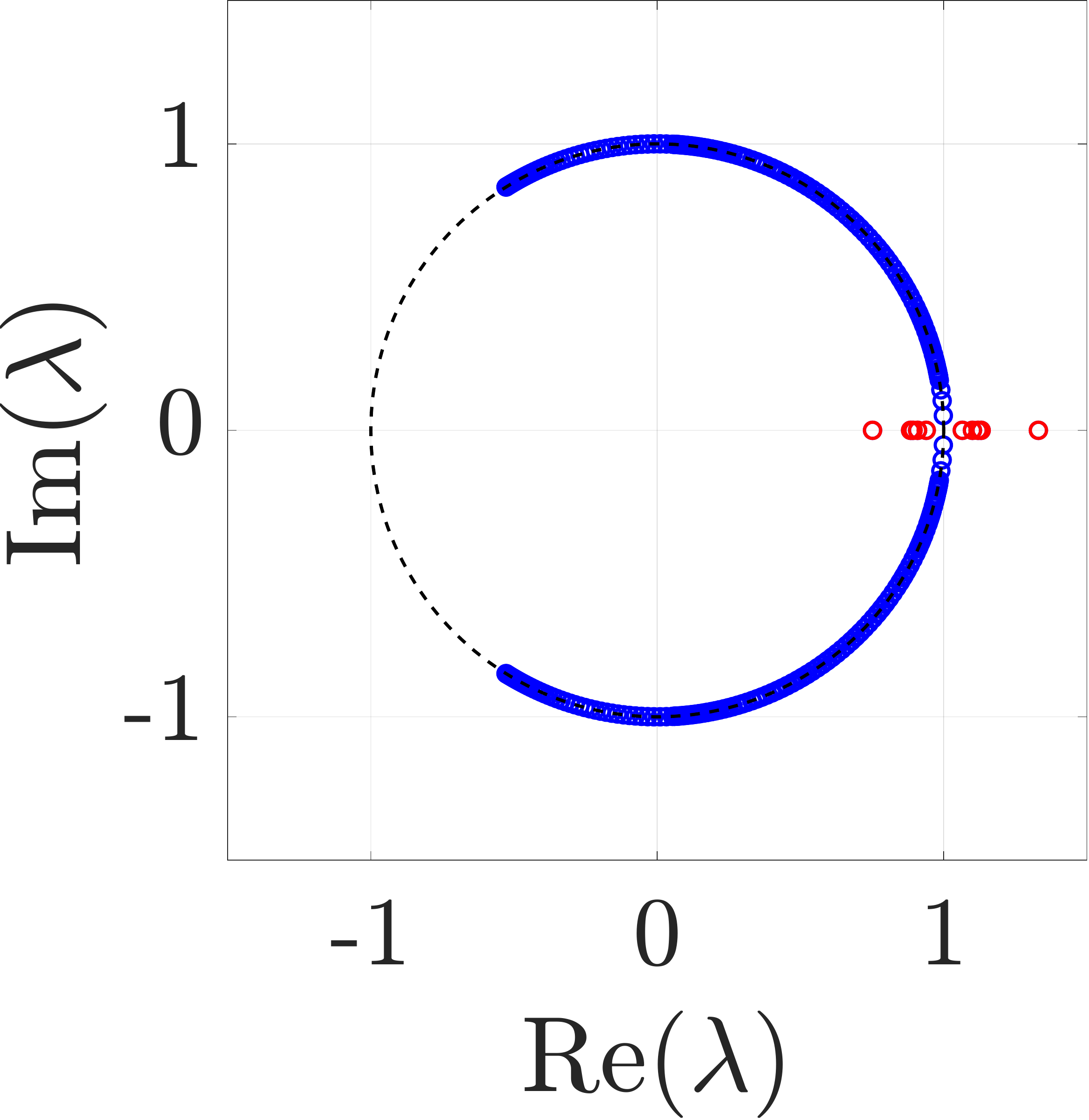}
		\label{fig:1D4a}}
	\caption{(Color online) The eigenvalues, $\lambda$ of the time evolution operator in Eq. \eqref{eq:SSQW-TimeEvolution} are plotted for $(\theta_1^1, \theta_2^1) = (-3 \pi/8, \pi/4)$ and $(\theta_1^0, \theta_2^0) = (-3 \pi/8, 5\pi/8)$ and different values of $\gamma$.  In \subref{fig:1D1a} $\gamma = 0$, \subref{fig:1D2a} $\gamma = 0.2$, \subref{fig:1D3a} $\gamma = $ min $(\gamma^1, \gamma^2) = 0.2110$, and \subref{fig:1D4a} $\gamma = 0.25$.}
	\label{fig:1DBE}
\end{figure*}

In Fig.~\ref{fig:1DBE}, we plot the complex eigenvalues $\lambda$ of the non-unitary evolution operator $U$ given by \eqref{eq:SSQW-TimeEvolution}. Here the close chain of length $201$ is divided into two parts of length $L = 100$ and $L = 101$ lattice sites. The boundaries are denoted by points $L_B = \pm50$. We have chosen $(\theta_1^1, \theta_2^1) = (-3 \pi/8, \pi/4)$ for $n > \abs{L_B}$ and $(\theta_1^0, \theta_2^0) = (-3 \pi/8, 5\pi/8)$ for  $n \le \abs{L_B}$ corresponding to winding numbers $W = 1$ and $W = 0$, respectively.

In Fig.~\ref{fig:1D1a}, we observe that two of the eigenvalues of the operator $U$ lying on the real axis, signifying the states with energy $0$ or $\pi$ for $\gamma = 0$. These states were absent in the homogeneous case; therefore, they are the edge states.  As we introduce scaling factor i.e. $\gamma \ne 0$,  the same behaviour persists until we reach the critical value of $\gamma$.  Since we have two sets of $\theta_1,~\theta_2$ which correspond to two different energy landscapes, we will have different exceptional points. Using Eq. \eqref{Eq:Delta-c}, these exceptional points come out to be $\gamma_c^1 = 0.2110$ and $\gamma_c^0 = 0.2832$ for the given choice of rotation parameters $\theta$'s. We find that the edge states persist till the point given by min($\gamma_c^0, \gamma_c^1$) after which we will have a complex spectrum for the Hamiltonian and we get many states with pure real $\lambda$ which have a contribution from broken exact $\mathcal{PT}$-symmetry.

In the case of non-Hermitian 2D DTQW, it is more difficult to establish the bulk-edge correspondence. This is mainly due to the fact that 2D DTQW does not support $\mathcal{PT}$-symmetry. The spectrum becomes complex as soon as we introduce the scaling which is evident from Eq. \eqref{eq: No PT-Symmetry}. In order to see the persistence of edge states, we only plot the real part of the eigenvalues of the Hamiltonian. In the case of 2D DTQW, we introduce the boundary by considering position-dependent coin operator only along the $y$-axis while keeping the $x$-direction periodic, as shown in Fig.~\ref{fig:BulkEdge2D}. For one part of the lattice, we choose $(\theta_1^{+1}, \theta_2^{+1}) = (7 \pi/6, 7 \pi/6)$ and for the other, we choose $(\theta_1^0, \theta_2^0) = (3 \pi/2, 2\pi/2)$; hence, the Chern numbers are  $C = +1$ and $0$ for the two parts.

\begin{figure*}
	\centering
	\subfigure[]{
		\includegraphics[width=4.1cm]{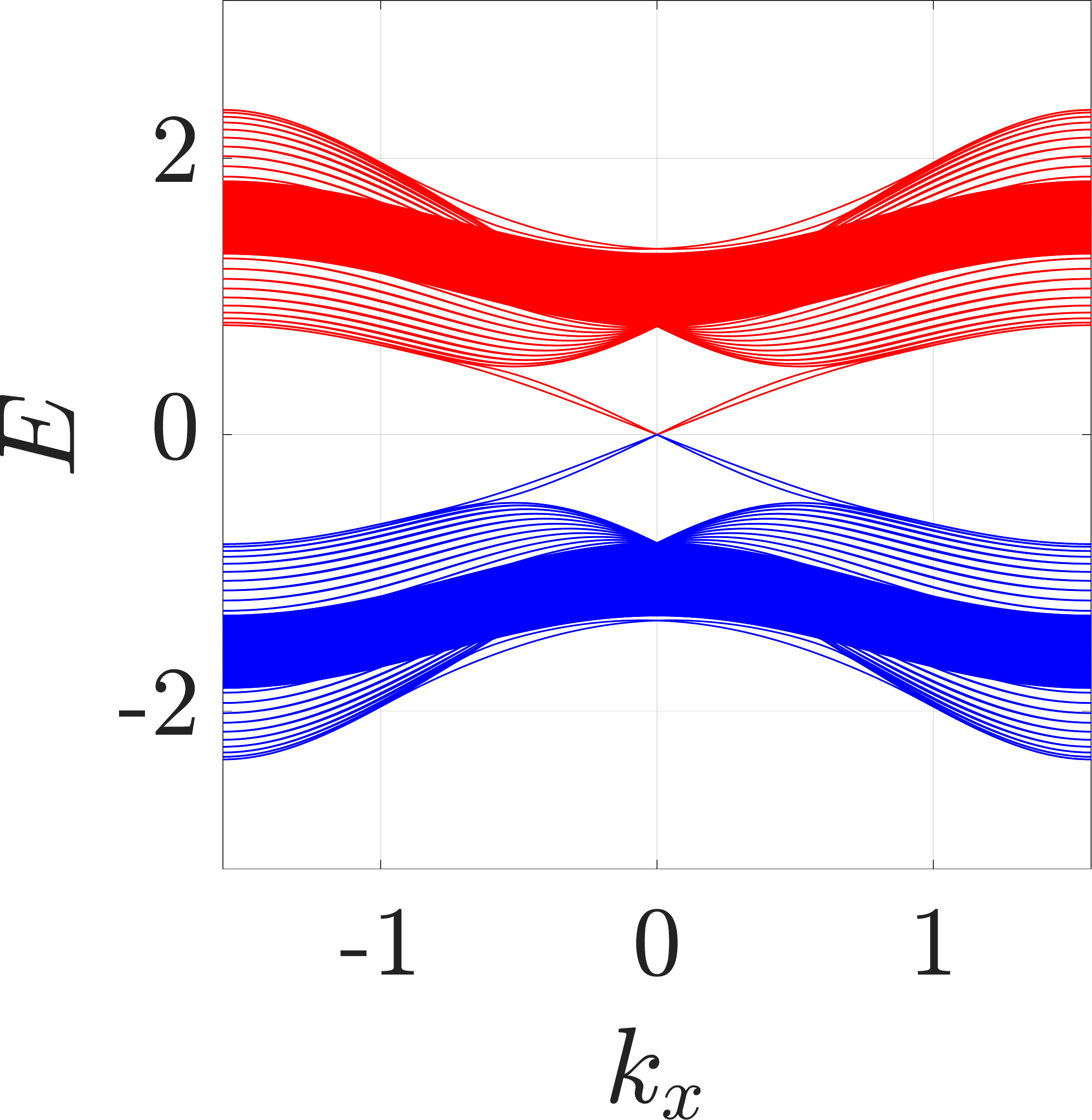}
		\label{fig:2D2a}}
	\subfigure[]{
		\includegraphics[width=4.1cm]{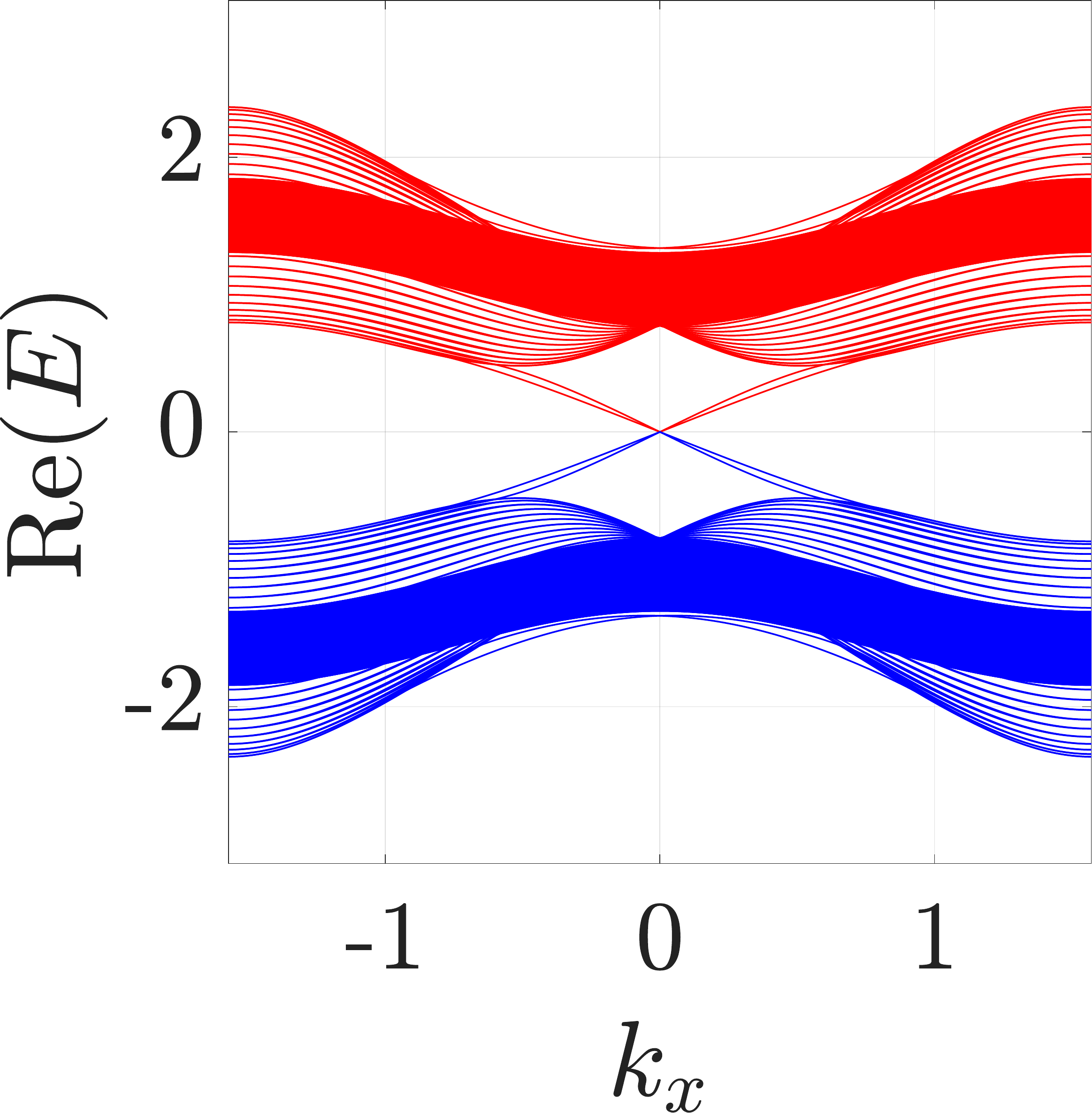}
		\label{fig:2D3a}}
	\subfigure[]{
		\includegraphics[width=4.1cm]{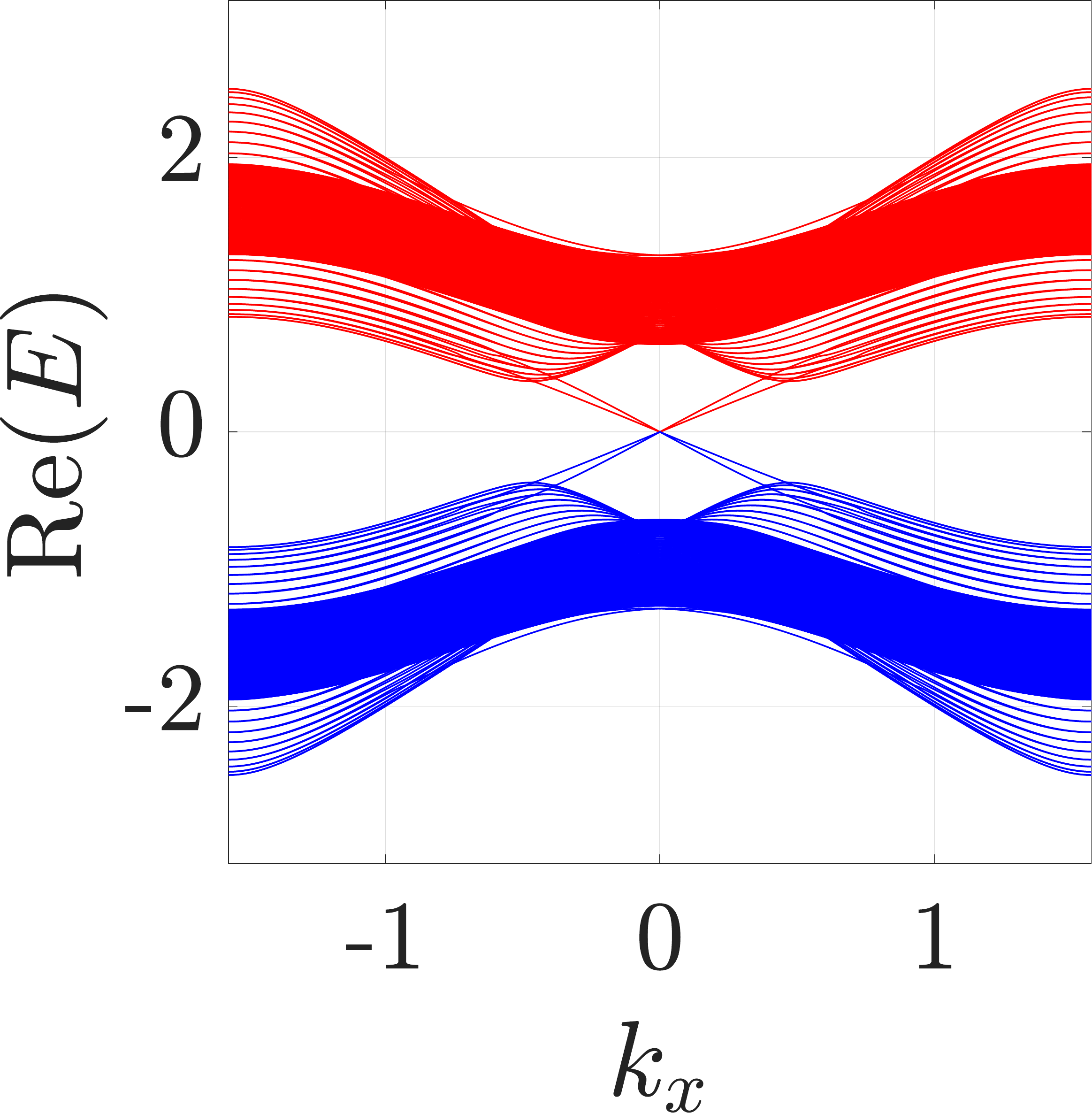}
		\label{fig:2D4a}}
	\subfigure[]{
		\includegraphics[width=4.1cm]{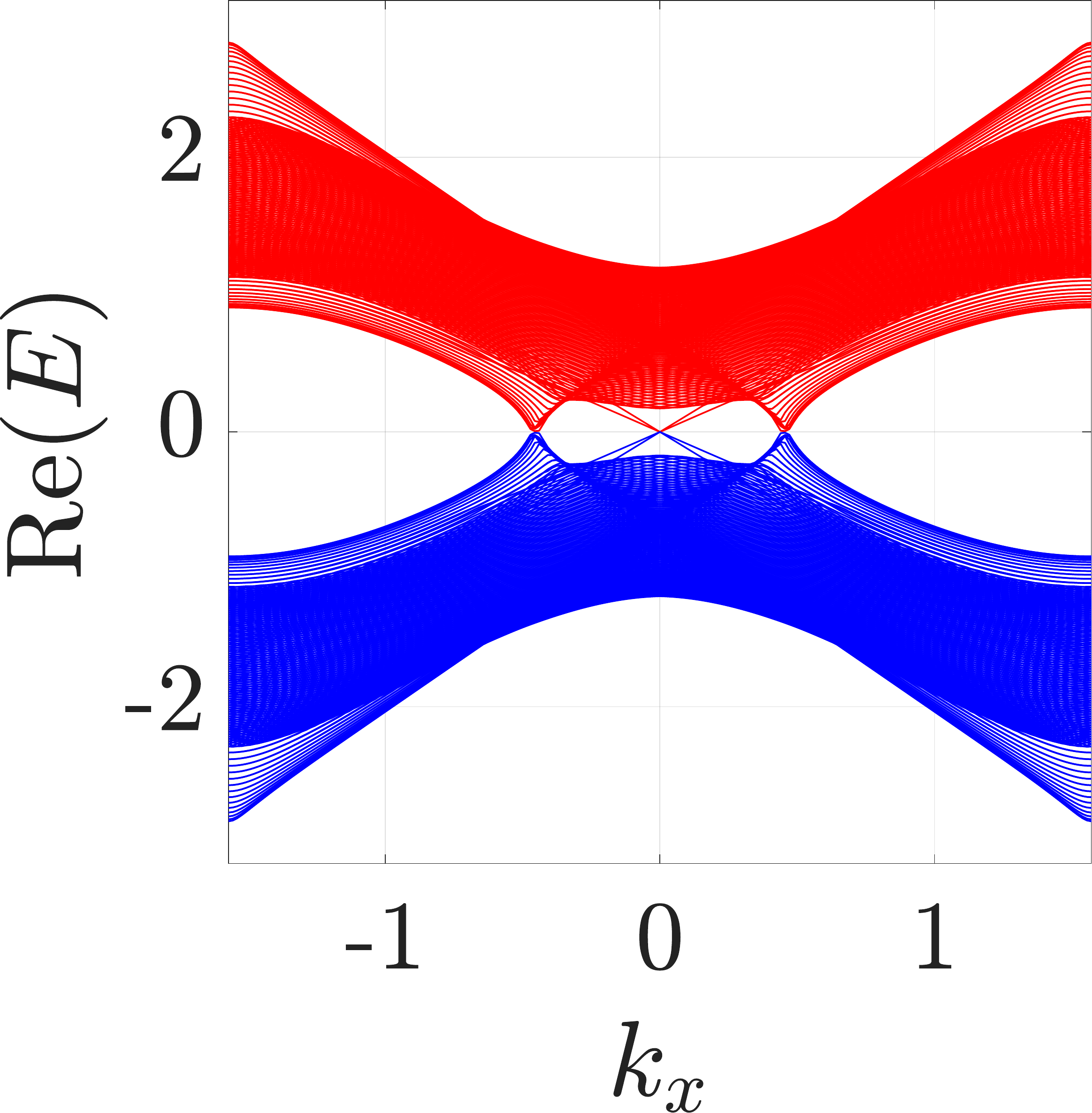}
		\label{fig:2D5a}}
	\caption{(Color online) Energy bands for the 2D DTQW are plotted for inhomogeneous lattice with lattice size $201 \times 201$. We have chosen  $(\theta_1^0, \theta_2^0) = (3 \pi/2, 2\pi/2)$ and $(\theta_1^{+1}, \theta_2^{+1}) = (7 \pi/6, 7\pi/6)$ which correspond to $C = 0$ and $C = +1$, respectively, for the two parts of the lattice. The scaling parameters  are chosen to be $\gamma_x =  \gamma_y = 0,~ 0.2,~0.3$ for \subref{fig:2D2a}, \subref{fig:2D3a} and \subref{fig:2D4a}, respectively. In all these figures we can see the edge states appearing on the boundaries of the two parts of the lattice. For larger values of the scaling parameter, i.e., $\gamma_x=\gamma_y = 0.47$ in Fig.~\subref{fig:2D5a} we see a large number of states between the two bands, which is due to the losses.}
	\label{fig:2DBE}
\end{figure*}

In Fig.~\ref{fig:2DBE}, we plot the real part of the spectrum as a function of the quasi-momentum in the $x$-direction. From these plots, we can see the persistence of the edge states even after introducing the scaling factor $\gamma$. For the large value of the scaling factors, we see the gap vanishes which is predominately due to the losses. Thus, it becomes very difficult to study the bulk-edge correspondence. 

\section{Conclusion}\label{Sec:Conclusion}
We have studied the effect of a lossy environment on the topological properties of discrete-time quantum walks. Specifically, we have studied the 1D SSQW and 2D DTQW and observed the persistence of topological phases  against losses in these systems. The loss is incorporated using the non-Hermitian Hamiltonian approach, where we include a scaling parameter $\gamma$ which characterizes the non-Hermiticity. 
We find a strong correspondence between the spontaneous exact $\mathcal{PT}$-symmetry breaking and the loss of topological order in 1D SSQW, i.e, the system retains its topological order for any value of $\gamma$, as long as the system respects the exact $\mathcal{PT}$-symmetry. Due to the absence of $\mathcal{PT}$-symmetry in 2D DTQW, we do not observe such correspondence in these systems. 
However, we observe loss-induced topological phase transition where we see that increasing the scaling parameter $\gamma$ may transfer the system from one non-trivial topological phase to another. We studied the bulk-boundary correspondence in 1D and 2D DTQW and observe the robustness of edge states against the losses. Our results confirm the robustness of topological properties of DTQWs and the role of losses in a topological phase transition.

%\bibliography{nuqw}

\section*{Acknowledgements}

S.\,K.\,G.\,acknowledges the financial support from SERB-DST (File No. ECR/2017/002404). S.~D.~acknowledges the support of research grant (DST/INSPIRE/04/2016/001391) from DST-INSPIRE, Govt.\,of India.  V.~M.~ thanks International Centre for Theoretical Sciences (ICTS) for the hospitality during the program - Geometric phase in Optics and Topological Matter (Code:  ICTS/geomtop2020/01).

\section*{Author contributions statement}
V.M. and S.K.G. conceived the idea and carried out the calculations. A.R. and S.D. contributed in developing the idea. All the authors contributed in writing the manuscript. 

\section*{Additional information}

\textbf{Competing interests:} The authors declare no competing interests.

\end{document}